\documentclass[11pt]{article}
\usepackage{jheppub}
\usepackage{bm}
\usepackage{dsfont}
\usepackage{tabu}
\usepackage{mathrsfs}
\usepackage{empheq}
\usepackage{amsmath,amsfonts}
\usepackage{verbatim}
\usepackage{bibentry}
\allowdisplaybreaks
\makeatletter
\def\@fpheader{\relax}
\def\@fpheader{~}
\makeatother
\newcommand{\numberthis}{\addtocounter{equation}{1}\tag{\theequation}}
\newcommand{\ack}[1]{{\textcolor{red}{\textbf{(#1)}}}}

\newcommand{\cN}{\mathcal{N}}

\newcommand{\cW}{\mathcal{W}}
\newcommand{\cA}[1]{\mathscr{A}^{[#1]}}
\newcommand{\cB}[1]{\mathscr{B}^{[#1]}}
\newcommand{\cC}[1]{\mathscr{C}^{[#1]}}
\newcommand{\cD}[1]{\mathscr{D}^{[#1]}}
\newcommand{\cE}[2]{\mathscr{E}^{[#1]}_{#2}}
\newcommand{\cPhi}[2]{\Phi^{[#1]}_{#2}}
\newcommand{\cPsi}[2]{\Psi^{[#1]}_{#2}}
\newcommand{\qJ}{Q_J}
\newcommand{\qL}{Q_L}
\newcommand{\qW}{Q_W}
\newcommand{\qA}{Q_A}
\newcommand{\qG}{Q_G}
\newcommand{\qH}{Q_H}
\newcommand{\qS}{Q_S}
\newcommand{\qT}{Q_T}
\newcommand{\lJ}{\lambda_J}
\newcommand{\lL}{\lambda_L}
\newcommand{\lW}{\lambda_W}
\newcommand{\lA}{\lambda_A}
\newcommand{\lG}{\lambda_G}
\newcommand{\lH}{\lambda_H}
\newcommand{\lS}{\lambda_S}
\newcommand{\lT}{\lambda_T}
\newcommand{\nJ}{\nu_J}
\newcommand{\nL}{\nu_L}
\newcommand{\nW}{\nu_W}
\newcommand{\nA}{\nu_A}
\newcommand{\nG}{\nu_G}
\newcommand{\nH}{\nu_H}
\newcommand{\nS}{\nu_S}
\newcommand{\nT}{\nu_T}
\newcommand{\lU}{\eta}
\newcommand{\lV}{\gamma}
\newcommand{\lGp}{\alpha^{-}}
\newcommand{\lGm}{\alpha^{+}}
\newcommand{\lGmp}{\alpha^{\pm}}
\newcommand{\lUp}{\beta^{-}}
\newcommand{\lUm}{\beta^{+}}
\newcommand{\lUpm}{\beta^{\mp}}
\newcommand{\lUmp}{\beta^{\pm}}
\newcommand{\mJ}{\mu_{1}}
\newcommand{\mT}{\mu_{2}}
\newcommand{\mW}{\mu_{3}}
\newcommand{\mV}{\tilde{\mu}_{2}}
\newcommand{\mGp}{\mu^{-}_{\frac{3}{2}}}
\newcommand{\mGm}{\mu^{+}_{\frac{3}{2}}}
\newcommand{\mGmp}{\mu^{\pm}_{\frac{3}{2}}}
\newcommand{\mUp}{\mu^{-}_{\frac{5}{2}}}
\newcommand{\mUm}{\mu^{+}_{\frac{5}{2}}}
\newcommand{\mUpm}{\mu^{\mp}_{\frac{5}{2}}}
\newcommand{\mUmp}{\mu^{\pm}_{\frac{5}{2}}}
\begin{document}
\title{A note on the $\mathcal{N}=2$ super-$\mathcal{W}_3$ holographic dictionary}
\author[a]{Alejandra Castro,}
\author[b]{Alberto Faraggi,}
\author[c]{Israel Osorio}
\affiliation[a]{Institute for Theoretical Physics, University of Amsterdam, Science Park 904, Postbus 94485,\\
1090 GL Amsterdam, The Netherlands}
\affiliation[b]{Departamento de Ciencias F\'isicas, Facultad de Ciencias Exactas, Universidad Andr\'es Bello,  \\ Sazi\'e 2212, Piso 7, Santiago, Chile.}
\affiliation[c]{Instituto de F\'isica, Pontificia Universidad Cat\'olica de Chile,  \\ Casilla 306, Santiago, Chile.}
\emailAdd{a.castro@uva.nl}
\emailAdd{alberto.faraggi@unab.cl}
\emailAdd{ijosorio@uc.cl}
\abstract{
This is a long-overdue companion paper to \cite{Banados:2015tft}. We study the relation between $sl(3|2)$ Chern-Simons supergravity on AdS$_3$ and two-dimensional CFT's with $\mathcal{N}=2$ super-$\mathcal{W}_3$ symmetry. Specifically, we carry out a complete analysis of asymptotic symmetries in a basis that makes the superconformal structure transparent, allowing us to establish the precise dictionary between currents and transformation parameters in the bulk and their boundary counterparts. We also discuss the incorporation of sources and display in full detail the corresponding holographic Ward identities. By imposing suitable hermiticity conditions on the CFT currents, we identify the superalgebra $su(2,1|1,1)$ as the appropriate real form of $sl(3|2)$ in Lorentzian signature. We take the opportunity to review some of the properties of the $\mathcal{N}=2$ super-$\cW_3$ conformal algebra, including its multiplet structure, OPE's and spectral flow invariance, correcting some minor typos present in the literature.
}
\maketitle
\newpage
\section{Introduction}
It is a well-known fact that higher spin algebras obtained from the Hamiltonian reduction of Kac-Moody current algebras can be realized in terms of flat connections in Drinfeld-Sokolov form \cite{Drinfeld:1984qv}. This fits the general scheme of higher spin dualities involving three-dimensional Chern-Simons theory and two-dimensional CFT's with $\cW$-symmetry, where some recent and notable examples are the relation between $sl(N)$ Chern-Simons theory and CFT's with $\cW_N$ symmetry \cite{Campoleoni:2010zq,Campoleoni:2011hg}, and, more generally, the duality between $hs[\lambda]$ Chern-Simons theory and  CFT's with $\cW_{\infty}[\lambda]$ symmetry \cite{Henneaux:2010xg,Gaberdiel:2011wb}. See \cite{Gaberdiel:2012uj} for a review on this subject.

One appeal of these higher spin dualities is that they serve as ``simple'' examples of holography: the abundance of symmetry opens an opportunity to build solvable and tractable instances of  the AdS/CFT correspondence. Higher spin dualities also allow us to introduce non-linear and non-geometrical features classically, via the Chern-Simons formulation of the gravitational side of the correspondence. These are features we expect to arise in string theory dualities, but are generically difficult to quantify. 

Our focus in this note is on supersymmetric examples of higher spin holography, with the specific goal of building a detailed dictionary between flat connections in Chern-Simons theory and the corresponding currents and sources in the dual supersymmetric $\cW$-algebra. As a complement to \cite{Banados:2015tft}, we study what it is perhaps the simplest three-dimensional higher spin supergravity, namely, Chern-Simons theory based on two copies of the $sl(3|2)$ gauge algebra.\footnote{See \cite{Henneaux:2015ywa} for an example with hypersymmetry.} As we prove in detail below, this theory is holographically dual to a CFT with $\cN=2$ super-$\cW_{3}$ symmetry, in the semiclassical (large $c$) limit. It is important to mention that the relation between $sl(3|2)$ connections and CFT's with $\cW_{(3|2)}$ symmetry was uncovered long ago in the context of Hamiltonian reduction of current algebras \cite{Lu:1991ux}. More recent studies from an AdS/CFT perspective can be found in \cite{Tan:2012xi,Datta:2012km,Peng:2012ae,Chen:2013oxa,Datta:2013qja}. However, the explicit form of the holographic dictionary between the basic Chern-Simons and CFT variables for this case has not been fully laid out so far. The original reference \cite{Lu:1991ux} included a map for the currents (although in a rather cumbersome basis), but it did not provide the corresponding map between transformation parameters, implying in practice that the relation between bulk and boundary sources is missing. By analyzing the CFT Ward identities, the authors in \cite{Datta:2013qja} produced a dictionary for both currents and sources, albeit only in the bosonic sector. In what follows we fill these gaps. We perform a complete analysis of the asymptotic symmetries of the $sl(3|2)$ Chern-Simons theory in a basis that makes the supersymmetries transparent and allows to make contact with the modern literature. In addition to the dictionary for the currents, we establish the holographic relation for the infinitesimal transformation parameters as well as for the sources, which is a key ingredient in the thermodynamic analysis of black hole solutions \cite{Banados:2015tft}. This enables us to reproduce the holographic Ward identities in full detail. Crucially, our treatment includes the fermionic sector. After adjusting for conventions, our results agree with \cite{Datta:2013qja} for the bosonic truncation.

Carrying out this rather technical analysis for a specific instance of a higher spin duality might seem superfluous. Nevertheless, there are two aspects that are worth recording and highlighting. The first is the distinction between Chern-Simons connections in Euclidean and Lorentzian signatures. Having constructed the complete dictionary in a basis that respects the $\cN=2$ supermultiplet structure, we can impose suitable hermiticity conditions on the CFT currents in order to identify the real form of the bulk $sl(3|2)$ gauge superalgebra that properly codifies this structure. While many entries of the holographic dictionary can be worked out in the Euclidean formalism, there are several features of the correspondence that are intrinsically Lorentzian. For example, in \cite{Banados:2015tft} it was crucial to single out the real form $su(2,1|1,1)$ to successfully construct Killing spinors in the bulk, and therefore identify BPS configurations with real values of the $U(1)$ R-charge. It also allowed us to identify the branches of solutions that have a physically meaningful thermodynamical interpretation in Lorentzian signature. More broadly, in the context of black hole physics several Lorentzian processes do not have a clear Euclidean counterpart. This has been manifest in the study of eternal black holes and multi-boundary solutions in Chern-Simons theory \cite{Castro:2016ehj,Henneaux:2019sjx,Cotler:2020ugk}. We therefore consider important to clearly identify the Lorentzian gauge algebra when studying these setups.     

The second aspect regards the original literature on the $\cW_{(3|2)}$ algebra. There are some minor typos in the OPE's and composite operators recorded originally in \cite{Romans:1991wi} which we correct in this note. We have verified that our OPE's are indeed correct via three independent methods, namely, by checking explicitly that all the Jacobi identities are satisfied, by complying with the spectral flow automorphism of the $\cN=2$ super-$\cW_3$ algebra, and by correctly matching with the semiclassical OPE's obtained in Chern-Simons theory. Although the results displayed here are specific for $\cW_{(3|2)}$, we present general explanations of our methods that can be easily implemented for other super $\cW$-algebras. 

The paper is organized as follows. In section \ref{Section W32} we provide a concise summary of the $\cN=2$ super-$\cW_{3}$ algebra and its main properties. After a brief review of higher spin supergravity on AdS$_3$, section \ref{Section Dictionary} is devoted to establishing the explicit holographic dictionary between the $sl(3|2)$ Chern-Simons theory and the dual CFT with $\cN=2$ super-$\cW_{3}$ symmetry, including currents, transformation parameters, sources and Ward identities, as well as the choice of real form of $sl(3|2)$ in Lorentzian signature. We end in section \ref{Section Conclusion} with some brief conclusions. In appendix \ref{app: quantum composites} we display the composite operators appearing in the $\cW_{(3|2)}$ algebra and their spectral flow transformations, and appendix \ref{app: sl(3|2)} collects our conventions on the superalgebra $sl(3|2)$ and its real form $su(2,1|1,1)\,$.
\section{The $\cN=2$ super-$\cW_{3}$ algebra}\label{Section W32}
We begin by reviewing some basic aspects of the $\cN=2$ super-$\cW_{3}$ algebra, often refered to as $\cW_{(3|2)}$. Far from being a comprehensive survey, this section is intended as a summary of the essential features that are pertinent for our purposes. In particular, we review the $\cN=2$ multiplet structure, the OPE's and the spectral flow invariance of the $\cW_{(3|2)}$ algebra. We also comment on the semiclassical limit needed to make contact with the holographic description and on the hermiticity properties of the generators on the cylinder. Important issues such as the commutator algebra and unitarity and BPS bounds are purposefully omitted (see \cite{Boucher:1986bh,Mizoguchi:1988vk,Romans:1991wi,Banados:2015tft}). The material below closely follows references \cite{Romans:1991wi,Candu:2012tr}.
\subsection{General structure and OPE's}
Let us briefly recall the structure of the $\cN=2$ super-Virasoro algebra. Besides the stress tensor $T$, this algebra contains a weight-1 $U(1)$ current $J$ and two weight-3/2 fermionic currents $G^+$ and $G^{-}$ with $U(1)$ charges $+1$ and $-1$, respectively. In standard conventions, their OPE's are given by
\begin{align}
	\label{Super Virasoro start}
	T(z)T(w)&\sim\frac{c/2}{(z-w)^{4}}+\frac{2T(w)}{(z-w)^{2}}+\frac{\partial T(w)}{z-w}\,,
	\\
	T(z)J(w)&\sim\frac{J(w)}{(z-w)^{2}} + \frac{\partial J(w)}{z-w}\,,
	\\
    T(z)G^{\pm}(w)&\sim\frac{3}{2}\frac{G^{\pm}(w)}{(z-w)^{2}}+\frac{\partial G^{\pm}(w)}{z-w}\,,
    \\
    J(z)J(w)&\sim\frac{c/3}{(z-w)^{2}}\,,
    \\
    J(z)G^{\pm}(w)&\sim\pm\frac{G^{\pm}(w)}{z-w}\,,
    \\
    \label{Super Virasoro finish}
   G^{+}(z)G^{-}(w)&\sim\frac{2c/3}{(z-w)^{3}}+\frac{2J(w)}{(z-w)^2}+\frac{2T(w)+\partial J(w)}{z-w}\,.
\end{align}
For notational convenience, we denote these fields collectively by $W^{(1)}=\left\{J,G^+,G^-,T\right\}$.
 
According to \cite{Candu:2012tr}, the $\cW$-algebras that are relevant for the $\cN=2$ version \cite{Creutzig:2011fe} of minimal model holography \cite{Gaberdiel:2010pz} contain, in addition to the super-Virasoro generators $W^{(1)}$, an infinite number of higher spin multiplets $W^{(s)}$, one for every integer spin $s\geq 2$. Each of these multiplets compromises four Virasoro primaries, $W^{(s)}=\left\{W^{(s)}_0,W^{(s)}_+,W^{(s)}_-,W^{(s)}_1\right\}$,\footnote{As usual, starting from the superconformal primary $W^{(s)}_{0}$, the other fields in the multiplet are generated by acting with $G^{\pm}$ as $W^{(s)}_{\pm}=G^{\pm}_{-\frac{1}{2}}W^{(s)}_{0}$ and $W^{(s)}_{1}=\frac{1}{4}\left(G^{+}_{-\frac{1}{2}}G^{-}_{-\frac{1}{2}}-G^{-}_{-\frac{1}{2}}G^{+}_{-\frac{1}{2}}\right)W^{(s)}_{0}$.} whose OPE's with the generators of the superconformal algebra are
\begin{align}
	\label{W1Ws OPE's start}
	T(z)W^{(s)}_0(w)&\sim \frac{sW^{(s)}_0(w)}{(z-w)^{2}}+\frac{\partial W^{(s)}_0(w)}{z-w}\,,
	\\
	T(z)W^{(s)}_1(w)&\sim\frac{(s+1)W^{(s)}_1(w)}{(z-w)^{2}}+\frac{\partial W^{(s)}_1(w)}{z-w}\,,
	\\
	T(z)W^{(s)}_{\pm}(w)&\sim\frac{\left(s+\frac{1}{2}\right)W^{(s)}_{\pm}(w)}{(z-w)^{2}}+\frac{\partial W^{(s)}_{\pm}(w)}{z-w}\,,
	\\
	J(z)W^{(s)}_1(w)&\sim\frac{sW^{(s)}_0(w)}{(z-w)^{2}}\,,
	\\
	J(z)W^{(s)}_{\pm}(w)&\sim\pm\frac{W^{(s)}_{\pm}(w)}{z-w}\,,
	\\
	G^{\pm}(z)W^{(s)}_0(w)&\sim\mp \frac{W^{(s)}_{\pm}(w)}{z-w}\,,
	\\
	G^{\pm}(z)W^{(s)}_1(w)&\sim\frac{\left(s+\frac{1}{2}\right)W^{(s)}_{\pm}(w)}{(z-w)^{2}}+\frac{1}{2}\frac{\partial W^{(s)}_{\pm}(w)}{z-w}\,,
	\\
	\label{W1Ws OPE's finish}
	G^{\pm}(z)W^{(s)}_{\mp}(w)&\sim\pm\frac{2sW^{(s)}_0(w)}{(z-w)^{2}}+\frac{2W^{(s)}_1(w)\pm\partial W^{(s)}_0(w)}{z-w}\,.
\end{align}
In particular, the conformal dimensions $\Delta$ and $U(1)$ charges $q$ of the different fields in the spin-$s$ multiplet $W^{(s)}$ are given in table \ref{multiplet}. 
\begin{table}[h]
\begin{center}
\begin{tabu}{c|[1.5pt]c|c}
\rule[-1ex]{0pt}{2.5ex} & $\Delta $ & $q$ \\ 
\tabucline[1.5pt]{-}\rule[-2ex]{0pt}{5ex} $W^{(s)}_{0}$ & $s$ & $\quad 0\quad $ \\ 
\hline 
\rule[-2ex]{0pt}{5ex} $W^{(s)}_{\pm}$ & $s+\frac{1}{2}$ & $\pm 1$ \\ 
\hline 
\rule[-2ex]{0pt}{5ex} $W^{(s)}_{1}$ & $s+1$ & $0$ \\ 
\end{tabu} 
\end{center}
\caption{Conformal dimensions and $U(1)$ charges of Virasoro primaries in an $\cN=2$ spin-$s$ multiplet.}
\label{multiplet}
\end{table}

As explained in \cite{Candu:2012tr,Gaberdiel:2012ku}, the Jacobi identities determine the full structure of the $\cN=2$ super-$\cW$ algebras up to two free parameters: the central charge $c$ and the self-coupling $c_{22,2}$ of the spin-2 primary $W^{(2)}_{0}$. Indeed, the singular part of the OPE of fields in the $W^{(2)}$ multiplet has the schematic form
\begin{align}
	W^{(2)}W^{(2)}&\sim n_{2}W^{(1)}+c_{22,2}W^{(2)}+c_{22,3}W^{(3)}\,.
\end{align}
Once a normalization for the currents is chosen, the parameters $n_{2}$ and $c_{22,3}$ are fixed in terms of $c$ and $c_{22,2}$, which are physically meaningful. The same is true for the structure constants appearing in the OPE's of all the other higher spin multiplets.

Quite interestingly, it possible to consistently decouple the multiplets $W^{(s)}$ with $s\geq3$ and truncate the full higher spin algebra to one containing just the super-Virasoro currents $W^{(1)}$ and the spin-2 multiplet $W^{(2)}$. The resulting algebra is precisely the $\cW_{(3|2)}$ algebra we want to study, and from now on we focus exclusively on this case. As discussed in \cite{Candu:2012tr}, the truncation happens when the parameter $c_{22,3}$ introduced above is zero, implying that the self-coupling $c_{22,2}$ is no longer independent but a particular function of the central charge. Fixing $n_{2} = c/2$ to comply with the standard normalization for the spin-2 current $W^{(2)}_0$, the decoupling occurs for
\begin{equation}\label{definition kappa}
	\frac{c_{22,2}}{2}=\pm \frac{(c+3)(5c-12)}{\sqrt{2(c+6)(c-1)(2c-3)(15-c)}}\equiv\kappa\,.
\end{equation}
Here we have introduced the constant $\kappa$ so as to make contact with the notation in Romans' paper \cite{Romans:1991wi}. As pointed out there, the sign ambiguity corresponds to the freedom of simultaneously flipping the sign of all fields in the $W^{(2)}$ multiplet. Notice that $\kappa$ is real only for $-6<c<1$ or $\frac{3}{2}<c<15$. In particular, it is purely imaginary as $c\to\infty$, which makes the representations of the algebra non-unitarity \cite{Romans:1991wi}.

In order to spell out the explicit form of the $\cN=2$ super-$\cW_3$ algebra we adopt the notation $W^{(2)}=\left\{V,U^{+},U^{-},W\right\}$ for the currents in the spin-2 multiplet. According to table \ref{multiplet}, $V$ has conformal dimension $2$ and $U(1)$ charge zero, $U^{\pm}$ have weight $5/2$ and $U(1)$ charge $\pm 1$, and $W$ has conformal dimension $3$ and $U(1)$ charge zero. The OPE's among these fields are given in \cite{Romans:1991wi} and read\footnote{Notice that $W(z)$ and $U^{\pm}(z)$ do not have the standard CFT normalization for a weight-$s$ current, namely,
\begin{equation*}
	J_{s}(z)J_{s}(w)\sim\frac{c/s}{(z-w)^{2s}}+\cdots\,.
\end{equation*}
Instead, we have adopted the same normalization as in \cite{Romans:1991wi} for ease of comparison. In particular, the currents $U^{\pm}$ and $W$ are related to their canonically-normalized counterparts by $U^{\pm}=\sqrt{5}U^{\pm}_{\text{canon}}$ and $W=\sqrt{\frac{15}{2}}W_{\text{canon}}$.}
\begin{align*}
	\numberthis\label{W2W2 OPE's start}
	V(z)V(w)&\sim\frac{c/2}{(z-w)^{4}}+\left(\frac{2}{(z-w)^{2}}+\frac{1}{z-w}\partial\right)\cA{2}(w)\,,
	\\
	W(z)W(w)&\sim\frac{5c/2}{(z-w)^{6}}+\left(\frac{2}{(z-w)^{2}}+\frac{1}{z-w}\partial\right)\cB{4}(w) 
	\\
	\numberthis
	&+\left(\frac{60}{(z-w)^{4}}+\frac{30}{(z-w)^{3}}\partial+\frac{9}{(z-w)^{2}}\partial^{2}+\frac{2}{z-w}\partial^{3} \right)\cB{2}(w)\,,
	\\
	\numberthis\label{VW OPE}
	V(z)W(w)&\sim\frac{\cC{4}(w)}{z-w}+\left(\frac{3}{(z-w)^{2}}+\frac{1}{z-w}\partial \right)\cC{3}(w)+\frac{36}{(z-w)^{4}}\cC{1}(w)\,,
	\\
	U^{+}(z)U^{-}(w)&\sim\frac{2c}{(z-w)^{5}}+\frac{\cD{4}(w)}{z-w}+\left(\frac{2}{(z-w)^{2}}+\frac{1}{z-w}\partial\right)\cD{3}(w)
	\\
	&+\left(\frac{20}{(z-w)^{3}}+\frac{10}{(z-w)^{2}}\partial+\frac{3}{z-w}\partial^{2} \right)\cD{2}(w)
	\\
	\numberthis
	&+\left(\frac{24}{(z-w)^{4}}+\frac{12}{(z-w)^{3}}\partial+\frac{4}{(z-w)^{2}}\partial^{2}+\frac{1}{z-w}\partial^{3} \right)\cD{1}(w)\,,
	\\
	\numberthis
	U^{\pm}(z)U^{\pm}(w)&\sim\frac{\cE{4}{\pm}(w)}{z-w}\,,
	\\
	V(z)U^{\pm}(w)&\sim\frac{\cPhi{7/2}{\pm}(w)}{z-w}+\left(\frac{5/2}{(z-w)^{2}}+\frac{1}{z-w}\partial \right)\cPhi{5/2}{\pm}(w)
	\\
	\numberthis
	&+\left(\frac{12}{(z-w)^{3}}+\frac{4 }{(z-w)^{2}}\partial+\frac{1}{z-w}\partial^{2} \right)\cPhi{3/2}{\pm}(w)\,,
	\\
	U^{\pm}(z)W(w)&\sim\frac{\cPsi{9/2}{\pm}(w)}{z-w}+\left(\frac{7/2}{(z-w)^{2}}+\frac{3/2}{z-w}\partial\right)\cPsi{7/2}{\pm}(w)
	\\
	&+\left(\frac{10}{(z-w)^{3}}+\frac{4}{(z-w)^{2}}\partial+\frac{1}{z-w}\partial^{2} \right)\cPsi{5/2}{\pm}(w)
	\\
	\numberthis\label{W2W2 OPE's finish}
	&+\left(\frac{60}{(z-w)^{4}}+\frac{20}{(z-w)^{3}}\partial +\frac{5}{(z-w)^{2}}\partial^{2}+\frac{1}{z-w}\partial^{3} \right)\cPsi{3/2}{\pm}(w)\,.
\end{align*}
The fields $\cA{s}$, $\cB{s}$, $\cC{s}$, $\cD{s}$, $\cE{s}{\pm}$, $\cPhi{s}{\pm}$, $\cPsi{s}{\pm}$ are built out of primary and quasi-primary composite operators, the precise form of which is fixed by the Jacobi identities \cite{Romans:1991wi}. We have reproduced them all in appendix \ref{app: quantum composites def}. For example, the operator appearing in the $VV$ OPE is
\begin{equation}
	\cA{2}=\frac{c}{c-1}\left(T-\frac{3}{2c}J^{2}\right)+\kappa V\,.
\end{equation}
Of course, normal ordering is assumed. The full $\cW_{(3|2)}$ OPE algebra is then given by equations \eqref{Super Virasoro start}-\eqref{Super Virasoro finish}, \eqref{W1Ws OPE's start}-\eqref{W1Ws OPE's finish} with $s=2$, and \eqref{W2W2 OPE's start}-\eqref{W2W2 OPE's finish}.

It is important to mention that we have detected some minor typos in the $G^{\pm}U^{\mp}$, $G^{\pm}W$ and $VU^{+}$ OPE's in reference \cite{Romans:1991wi}, which have been corrected in the expressions above. More importantly, however, \cite{Romans:1991wi} has typos in the definitions of the composite operators $\cC{1}$ and $\cPhi{7/2}{\pm}$ entering in the $VW$ and $VU^{\pm}$ OPE's.\footnote{The typo in $\cC{1}$ was already noted in \cite{Datta:2013qja}.} These are highlighted in \eqref{quantum composite C1} and \eqref{quantum composite Phi7/2}. The correct expressions can be found by checking that all the Jacobi identities are in fact satisfied, a task that we have performed exhaustively using the Mathematica package\footnote{The author kindly shared with us an updated version of the package.} introduced in \cite{Thielemans:1994er}. These corrections are also required by the spectral flow invariance of the algebra, as we explain next. 
\subsection{Spectral flow}\label{subsec:spectral}
It is a well-known fact that the $\cN=2$ superconformal algebra enjoys a continuous class of automorphisms called spectral flow. Although usually expressed in terms of modes, we find it more convenient for our purposes to write this symmetry directly in terms of the currents. It is easy to check that the transformations
\begin{empheq}{alignat=7}\label{spectral flow currents W1}
	T'(z)&=T(z)+\frac{\eta}{z} J(z)+\frac{c\eta^{2}}{6z^{2}}\,,
	&\qquad
	J'(z) &=J(z)+\frac{c\eta}{3z}\,,
	&\qquad
	G^{\pm'}(z)&=z^{\pm\eta}G^{\pm}(z)\,,
\end{empheq}
where $\eta$ is a continuous parameter, leave the OPE's \eqref{W1Ws OPE's start}-\eqref{W1Ws OPE's finish} invariant. In particular, for $\eta\in\mathds{Z}+\frac{1}{2}$ this operation interpolates between the NS sector and the Ramond sector, whereas for $\eta\in\mathds{Z}$ it maps each sector onto itself. 

As pointed out in \cite{Romans:1991wi}, the extension of spectral flow to the $\cW_{(3|2)}$ case is achieved by letting the spin-2 multiplet currents transform as
\begin{empheq}{alignat=7}\label{spectral flow currents W2}
	V'(z)&=V(z)\,,
	&\qquad
	W'(z)&=W(z)+\frac{2\eta}{z}V(z)\,,
	&\qquad
	U^{\pm'}(z)&=z^{\pm\eta}U^{\pm}(z)\,.
\end{empheq}
Using these rules it is straightforward, albeit tedious, to compute how of the various composite fields $\cA{s}$, $\cB{s}$, $\cC{s}$, $\cD{s}$, $\cE{s}{\pm}$, $\cPhi{s}{\pm}$, $\cPsi{s}{\pm}$ change. The results are written appendix \ref{app: quantum composites spectral flow}, together with an explanation of some of the subtleties involved in the calculation. Using the Mathematica package of \cite{Thielemans:1994er}, we have thoroughly checked that \eqref{spectral flow currents W1} and \eqref{spectral flow currents W2} are in fact a symmetry of the full $\cN=2$ super-$\cW_3$ OPE algebra. As with the Jacobi identities, this is true only if the composite operators $\cC{1}$ and $\cPhi{7/2}{\pm}$ of \cite{Romans:1991wi} are modified as shown in appendix \ref{app: quantum composites def}. 
 For example, spectral flow invariance of the $VW$ OPE \eqref{VW OPE} requires that the field $\cC{1}$ transform as (see appendix \ref{app: quantum composites spectral flow})
\begin{empheq}{alignat=7}
	\cC{1}\hspace{0.01cm}'(z)&=\cC{1}(z)+\frac{c\eta}{36z}\,.
\end{empheq}
This property follows from \eqref{spectral flow currents W1} only if we use the correct coefficient in \eqref{quantum composite C1}. A similar approach can be taken to fix the composite operator $\cPhi{7/2}{\pm}$ appearing in the $VU^{\pm}$ OPE (cf. \eqref{quantum composite Phi7/2}).

The spectral flow automorphism of the $\cW_{(3|2)}$ algebra will be an important guiding principle in the upcoming discussion of the holographic dictionary. For the time being, we anticipate that some of the bulk fields turn out to be spectral flow-invariant, so their CFT duals must posses the same property. Here we construct two such quantities. Using \eqref{spectral flow currents W1} and \eqref{spectral flow currents W2}, it is easy to see that (normal ordering is implicit)
\begin{empheq}{alignat=7}\label{spectral flow invariant operators}
	T'-\frac{3}{2c}J'J'=T-\frac{3}{2c}JJ\,,
	&\qquad
	W'-\frac{6}{c}J'V' =W-\frac{6}{c}JV\,,
\end{empheq}
leading to two invariant quasi-primary operators of dimensions 2 and 3, respectively. As we will see below, these particular combinations of fields appear naturally from the bulk perspective. Coincidentally, it follows from the first relation in \eqref{spectral flow invariant operators} that ${\cA{2}}'=\cA{2}$, resulting in the invariance of the $VV$ OPE. The fact that the composite $\cA{2}$ does not change under spectral flow makes this particular check fairly simple. Proving the invariance of the remaning OPE's, however, is more involved.
\subsection{Semiclassical limit}\label{subsec: semiclassical}
In order make contact with the holographic description in the next section we need to extract the semiclassical limit of the OPE relations \eqref{Super Virasoro start}-\eqref{Super Virasoro finish}, \eqref{W1Ws OPE's start}-\eqref{W1Ws OPE's finish} and \eqref{W2W2 OPE's start}-\eqref{W2W2 OPE's finish}, which in the bulk Chern-Simons theory translate into classical Poisson brackets. This involves taking a ``large-$c$" and ``large-current" limit, procedure that is more subtle than a naive expansion in $1/c$. As outlined in \cite{Candu:2013uya}, the proper way to proceed is to rescale all the CFT currents as
\begin{empheq}{alignat=7}
	J_i(z)&=c\,\tilde{J}_i(z)
\end{empheq}
and then expand for $c\to\infty$ while keeping $\tilde{J}_i$ fixed. One finds that the leading term in the $\tilde{J}\tilde{J}$ OPE's is of order $1/c$, that is,
\begin{empheq}{alignat=7}
	\tilde{J}_i(z)\tilde{J}_j(w)&=\frac{1}{c}\tilde{J}_i(z)\tilde{J}_j(w)\Big|_{\textrm{semiclass}}+O\left(1/c^2\right)\,.
\end{empheq}
Notice that by construction the OPE's on the right hand side do not depend explicitly on the central charge. We can now express everything back in terms of the original currents $J_i$ and write
\begin{empheq}{alignat=7}
	J_i(z)J_j(w)\Big|_{\textrm{semiclass}}&=c\tilde{J}_i(z)\tilde{J}_j(w)\Big|_{\textrm{semiclass}}\,.
\end{empheq}
This defines the semiclassical limit of the algebra.

Following the above procedure, we find that the semiclassical version of the $\cN=2$ super-$\cW_3$ algebra is identical to its quantum progenitor \eqref{Super Virasoro start}-\eqref{Super Virasoro finish}, \eqref{W1Ws OPE's start}-\eqref{W1Ws OPE's finish} and \eqref{W2W2 OPE's start}-\eqref{W2W2 OPE's finish} with the proviso that we use the semiclassical limit of the composites $\cA{s}$, $\cB{s}$, $\cC{s}$, $\cD{s}$, $\cE{s}{\pm}$, $\cPhi{s}{\pm}$, $\cPsi{s}{\pm}$, as opposed to their full quantum expressions \eqref{quantum composites start}-\eqref{quantum composites finish}. For example, the $VV$ OPE is still given by
\begin{empheq}{alignat=7}
	V(z)V(w)\Big|_{\textrm{semiclass}}&\sim\frac{c/2}{(z-w)^{4}}+\left(\frac{2}{(z-w)^{2}}+\frac{1}{z-w}\partial\right)\cA{2}(w)\,,
\end{empheq}
but with
\begin{empheq}{alignat=7}
	\cA{2}&\xrightarrow[\text{semiclass}]{}T-\frac{3}{2c}J^2+\kappa V\,,
	&\qquad
	\kappa&\xrightarrow[\text{semiclass}]{}\pm\frac{5i}{2}\,.
\end{empheq}
The full list of semiclassical composite fields is written in the next section. As we will see, they are in perfect agreement with the corresponding expressions obtained from the bulk analysis.
\subsection{From the plane to the cylinder}\label{subsec:hermite}
In the final portion of this section we collect some useful relations regarding the hermiticity and spectral flow properties of the CFT currents cast on the Euclidean cylinder versus the complex plane. Since bulk observables are naturally defined on the AdS$_3$ cylinder, this will ease the comparison with the dual Chern-Simons description, especially when discussing the continuation of the higher spin theory from Euclidean to Lorentzian signature and identifying the correct real form of the $sl(3|2)$ gauge algebra.

Recall that in radial quantization a real quasi-primary field of dimensions $(h,\bar{h})$ satisfies the hermiticity condition
\begin{empheq}{alignat=7}\label{conjugation plane}
	\Phi_{\textrm{plane}}(z,\bar{z})^{\dagger}&=\bar{z}^{-2h}z^{-2\bar{h}}\Phi_{\textrm{plane}}\left(\frac{1}{\bar{z}},\frac{1}{z}\right)\,.
\end{empheq}
Similarly, for a pair of charge conjugate fields this is
\begin{empheq}{alignat=7}
	\label{conjugation plane charged}
	\Phi^{\pm}_{\textrm{plane}}(z,\bar{z})^{\dagger}&=\bar{z}^{-2h}z^{-2\bar{h}}\Phi^{\mp}_{\textrm{plane}}\left(\frac{1}{\bar{z}},\frac{1}{z}\right)\,.
\end{empheq}
One can check that in the conventions adopted here the $\cW_{(3|2)}$ algebra is consistent with these conditions when imposed on the genetators $\left\{J,T,\kappa V,\kappa W\right\}$ and $\left\{G^{\pm},\kappa U^{\pm}\right\}$. Importantly, as required by the OPE structure, the currents in the spin-3 multiplet must always be accompanied by the coupling $\kappa$, which is imaginary in the semiclassical limit \cite{Banados:2015tft}.

The labels in \eqref{conjugation plane} and \eqref{conjugation plane charged} emphasize the fact that these relations are valid on the complex plane, where the OPE's are defined. The transition from the plane to the cylinder is achieved through the conformal transformation $z\to e^{\zeta z}$, where $\zeta$ is a bookkeeping device that allows us to accommodate different conventions relating the real cylinder coordinates $-\infty<t_E<\infty$ and $\phi\sim\phi+2\pi$ to the complex pair $z=\zeta^{-1}\left(t_E+i\phi\right)$ and $\bar{z}=\bar{\zeta}^{-1}\left(t_E-i\phi\right)$. Then, taking into account the conformal weights of the fields, the hermiticity conditions \eqref{conjugation plane} and \eqref{conjugation plane charged} become, respectively,
\begin{empheq}{alignat=7}\label{conjugation cylinder}
	\Phi_{\textrm{cyl}}(z,\bar{z})^{\dagger}&=\left(\frac{\bar{\zeta}}{\zeta}\right)^{h-\bar{h}}\Phi_{\textrm{cyl}}\left(-\frac{\bar{\zeta}}{\zeta}\bar{z},-\frac{\zeta}{\bar{\zeta}}z\right)\,,
\end{empheq}
and
\begin{empheq}{alignat=7}
	\label{conjugation cylinder charged}
	\Phi^{\pm}_{\textrm{cyl}}(z,\bar{z})^{\dagger}&=\left(\frac{\bar{\zeta}}{\zeta}\right)^{h-\bar{h}}\Phi^{\mp}_{\textrm{cyl}}\left(-\frac{\bar{\zeta}}{\zeta}\bar{z},-\frac{\zeta}{\bar{\zeta}}z\right)\,.
\end{empheq}
Ultimately, the different phases can be understood by recalling that in the Euclidean formalism the effect of complex conjugation on the time direction $t=-it_E$ must be compensated in the definition of Hermitian conjugate by taking $t_E\to-t_E$.\footnote{In terms of the mode expansions (setting $\bar{h}=0$ for simplicity)
\begin{empheq}{alignat*=7}
	\Phi_{\textrm{plane}}(z)&=\sum_{n}\frac{\Phi_n}{z^{n+h}}\,,
	&\qquad
	\Phi_{\textrm{cyl}}(z)&=\zeta^h\sum_{n}\Phi_ne^{-n\zeta z}\,,
\end{empheq}
both Hermiticity conditions imply that $\Phi_n^{\dagger}=\Phi_{-n}$.}

Lastly, we spell out the spectral flow transformations for the CFT currents on the cylinder. Upon implementing the conformal map $z\to e^{\zeta z}$, one readily finds that relations \eqref{spectral flow currents W1} and \eqref{spectral flow currents W2} become
\begin{equation}\label{spectral flow cylinder currents W1}
\begin{aligned}
	T'_{\textrm{cyl}}(z)&=T_{\textrm{cyl}}(z)+\zeta\eta J_{\textrm{cyl}}(z)+\frac{\zeta^2c\eta^2}{6}\,,
	\\
	J'_{\textrm{cyl}}(z)&=J_{\textrm{cyl}}(z)+\frac{\zeta c\eta}{3}\,,
	\\
	G^{\pm'}_{\textrm{cyl}}(z)&=e^{\pm\eta \zeta z}G^{\pm}_{\textrm{cyl}}(z)\,,
\end{aligned}
\end{equation}
and
\begin{equation}\label{spectral flow cylinder currents W2}
\begin{aligned}
	V'_{\textrm{cyl}}(z)&=V_{\textrm{cyl}}(z)\,,
	\\
	W'_{\textrm{cyl}}(z)&=W_{\textrm{cyl}}(z)+2\zeta\eta V_{\textrm{cyl}}(z)\,,
	\\
	U^{\pm'}_{\textrm{cyl}}(z)&=e^{\pm\eta \zeta z}U^{\pm}_{\textrm{cyl}}(z)\,.
\end{aligned}
\end{equation}
Of course, the operators appearing in \eqref{spectral flow invariant operators} remain spectral flow-invariant. This version of the transformations is better suited for comparison with the analogous relations appearing in the bulk Chern-Simons theory.
\section{The $\cN=2$ super-$\cW_{3}$ holographic dictionary}\label{Section Dictionary}
Having reviewed some of the properties of the $\cN=2$ super-$\cW_{3}$ algebra, we now move on to study its realization in terms of higher spin fields on AdS$_3$. We first provide a short summary of Chern-Simons supergravity theory, touching only on those points that are relevant to the construction of the holographic dictionary. Subsequently, a detailed derivation of the asymptotic symmetry algebra is given for the $sl(3|2)$ case, followed by a discussion of sources and the corresponding holographic Ward identities, which are relevant for the study of higher spin black hole solutions. The correct choice of real form in Lorentzian signature is also discussed. Our conventions for $sl(3|2)$ follow \cite{Banados:2015tft} and are reproduced in appendix \ref{app: sl(3|2)} for completeness. Since this superalgebra has dimension $24$ and involves $5\times5$ matrices, we have found it necessary to use a mathematical software such as Maple and Mathematica to perform most of the calculations.
\subsection{Higher spin supergravity on AdS$_3$}\label{subsec:HSreview}
In its simplest version, the action for three-dimensional higher spin gravity with negative cosmological constant is\footnote{We refer the reader to the extensive literature for a more detailed review; see for example \cite{Campoleoni:2011hg, Gaberdiel:2012uj,Ammon:2012wc, Castro:2016tlm} and references therein.} 
\begin{empheq}{alignat=7}
	I_{CS}&=\frac{k_{cs}}{4\pi}\int_M\textrm{Tr}\Big[CS(A)-CS(\bar{A})\Big]\,,
	&\qquad
	CS\left(A\right)&=A\wedge dA+\frac{2}{3}A\wedge A\wedge A\,,
\end{empheq}
where $A$ and $\bar{A}$ are two independent connections valued in a real Lie algebra $\mathfrak{g}$ and $\textrm{Tr}$ denotes the trace in some representation of choice.  As appropriate to AdS$_3$, the topology of the spacetime $M$ is assumed to be that of a solid cylinder, with coordinates $(\rho,t,\phi)$ such that $-\infty< t<\infty$, $\phi\sim\phi+2\pi$ and the boundary is located at $\rho\to\infty$. The corresponding equations of motion read
\begin{empheq}{alignat=7}
	dA+A\wedge A&=0\,,
	&\qquad
	d\bar{A}+\bar{A}\wedge\bar{A}&=0\,,
\end{empheq}
implying that both connections are flat. Local symmetries include diffeomorphisims and gauge transformations 
\begin{empheq}{alignat=7}\label{gauge transformations general}
	\delta A&=d\Lambda+\left[A,\Lambda\right]\,,
	&\qquad
	\delta\bar{A}&=d\bar{\Lambda}+\left[\bar{A},\bar{\Lambda}\right]\,,
	&\qquad
	\Lambda,\bar{\Lambda}&\in\mathfrak{g}\,,
\end{empheq}
although the former can be seen as a particular case of the latter when on-shell. The gravitational sector of the theory is associated with an $sl(2;\mathds{R})$ subalgebra, whose generators $L_i$ satisfy\footnote{The constant $k_{cs}$ appearing in the action is related to Newton's constant $G_3$ and the AdS$_3$ radius $l$ by ${\displaystyle k_{cs}=\frac{l}{8G_3\textrm{Tr}\left[L_0^2\right]}}$. The Chern-Simons level of the $sl(2)$ gravitational theory is ${\displaystyle k=\frac{l}{4G_3}}$.}
\begin{empheq}{alignat=7}
	[L_i,L_j]&=(i-j)L_{i+j}\,,
	&\qquad
	L_i&\in\mathfrak{g}\,.
\end{empheq}
The spectrum of higher spin fields then depends on the precise way in which this subalgebra is embedded in $\mathfrak{g}$; different embeddings give rise to different field contents in the bulk.   A supersymmetric extension can be obtained by considering instead a Lie superalgebra and replacing the trace $\textrm{Tr}$ by the supertrace $\textrm{sTr}$. In this case one must specify the embedding of $osp(1|2)\supset sl(2)$ in the gauge superalgebra.

Two-dimensional CFT's are usually discussed in Euclidean signature, so is convenient to also formulate the supergravity theory in this language. Our conventions follow \cite{deBoer:2013gz,deBoer:2014fra} and are such that after the Wick rotation $t=-it_E$, the light-cone directions $x^{\pm}=t\pm\phi$ become complex coordinates $x^+\to z$ and $x^-\to-\bar{z}$, subject to the periodicity conditions $z\sim z+2\pi$ and $\bar{z}\sim\bar{z}+2\pi$. Depending on the type of solutions one is interested in, one can generalize this condition and let $(z,\bar{z})$ parametrize any Riemann surface. The cylinder is the topology most fitting for the analysis of asymptotic symmetries with AdS$_3$ boundary conditions, whereas the torus $(z\sim z+2\pi\sim z+2\pi\tau)$ is appropriate for the discussion of black hole solutions \cite{Gutperle:2011kf,Kraus:2011ds,Gaberdiel:2012yb,Banados:2012ue,deBoer:2013gz,Ammon:2012wc,Bunster:2014mua,Perez:2013xi,Chen:2013oxa,Datta:2013qja,Henneaux:2013dra,Perez:2012cf,Banados:2015tft,Castro:2016tlm,Banados:2016hze}. In this paper, will be concerned exclusively with the former. Notice that in these conventions the map between the cylinder and the plane is $z\to e^{iz}$.

As it turns out, when continuing to imaginary time, the algebra $\mathfrak{g}$ needs to be complexified and the two connections $A$ and $\bar{A}$ are no longer independent. Rather, they are related by
\begin{empheq}{alignat=7}\label{Euclidean relation}
	\bar{A}&=-A^{\dagger}\,,
\end{empheq}
a condition that ensures the reality of the action and of all other physical observables. Recall, however, that a complex algebra can have several real forms, so special care must be taken in order to reconstruct the appropriate Lorentzian theory. We will come back to this point below. From now on we focus on the unbarred sector only.

Boundary conditions are a crucial ingredient in the context of the AdS/CFT correspondence. As shown in \cite{Campoleoni:2010zq}, using the gauge symmetries \eqref{gauge transformations general} of the Chern-Simons theory, we can eliminate the radial dependence of the connection and write
\begin{empheq}{alignat=7}
	A(\rho,z,\bar{z})&=b^{-1}(\rho)\Big(a(z,\bar{z})+d\Big)b(\rho)\,,
\end{empheq}
for some suitable choice of group element $b(\rho)$. This allows us to work only with the two-dimensional boundary connection $a(z,\bar{z})$, whose components satisfy the equation
\begin{empheq}{alignat=7}\label{flatness}
	\partial a_{\bar{z}}-\bar{\partial}a_z+[a_z,a_{\bar{z}}]&=0\,.
\end{empheq}
As further argued in \cite{Campoleoni:2010zq,Campoleoni:2011hg,Henneaux:2010xg,Gaberdiel:2011wb}, by a series of gauge transformations, any asymptotically AdS connection can be brought to the so-called highest weight gauge, or Drinfeld-Sokolov form, 
\begin{empheq}{alignat=7}\label{highest-weight}
	a_z&=L_1+\bm Q(z)\,,
	&\qquad
	a_{\bar{z}}&=0\,,
\end{empheq}
where $\bm Q\in\mathfrak{g}$ is a matrix satisfying $[L_{-1},\bm Q]=0$. The holomorphicity of $a_z$ follows from the flatness condition \eqref{flatness}. These represent {\it source-free} solutions. In particular, the connection corresponding to pure AdS$_3$ has $\bm Q=\frac{1}{4}L_{-1}$. Other source-free configurations include boundary gravitons (higher spin generalizations of Brown-Henneaux states \cite{Brown:1986ed}) and conical defect solutions \cite{Castro:2011iw}. The deformation of AdS boundary conditions by the incorporation of sources will be discussed in section \ref{subsec:sources}.
\subsection{Dictionary part I: currents and asymptotic symmetries}\label{subsec: asymp symm}
The asymptotic symmetries of the higher spin theory with AdS boundary conditions are defined as those residual gauge transformations \eqref{gauge transformations general} that preserve the form of the Drinfeld-Sokolov connection \eqref{highest-weight}. Concretely, one looks for gauge parameters $\Lambda$ such that
\begin{empheq}{alignat=7}\label{DS variation general}
	\partial\Lambda+\left[L_1+\bm Q,\Lambda\right]&=\delta\bm Q\,,
	&\qquad
	[L_{-1},\delta\bm Q]&=0\,.
\end{empheq}
Since $\Lambda$ must be holomorphic in order for the transformation to be compatible with $a_{\bar{z}}=0$, we can expand it as
\begin{empheq}{alignat=7}\label{gauge parameter general}
	\Lambda&=\bm\lambda(z)+\cdots\,,
\end{empheq}
where $[L_1,\bm\lambda]=0$, i.e. a lowest-weight condition. The dots represent higher-weight terms that are fixed algebraically in terms of $\bm Q$, $\bm\lambda$ and their derivatives by the condition \eqref{DS variation general}. Naturally, $\bm Q$ is allowed to change under the asymptotic symmetry transformations, this being precisely the algebra one is interested in uncovering. According to the AdS/CFT correspondence, the different components of $\bm Q(z)$ and $\bm\lambda(z)$ are then identified, respectively, with the currents $J_s(z)$ and parameters $\epsilon_s(z)$ generating the same symmetry algebra in a dual CFT. We will now go through some of the details of this analysis in the case of Chern-Simons supergravity based on the superalgebra $\mathfrak{g}=sl(3|2)$ and show that the resulting structure is given by the semiclassical limit of the $\cN=2$ super-$\cW_3$ algebra reviewed in section \ref{Section W32}.
\subsubsection{Spectrum and operator content}
The first step in the construction of the holographic dictionary is to derive the spectrum of fields in the bulk and verify that it matches the operator content of the dual CFT. To this purpose, we look at the decomposition of the adjoint representation of $sl(3|2)$ into irreducible representations of $osp(1|2)$. Concretely, we focus our attention on the principal embedding of $osp(1|2)$ in $sl(3|2)$, since this is the case that makes contact with the $\cW_{(3|2)}$ algebra. The decomposition reads (see e.g. \cite{Peng:2012ae})
\begin{empheq}{alignat=7}\label{Adjoint decomposition}
	\text{adj}\left[sl(3|2)\right]&=\mathcal{R}_{1/2}\oplus\mathcal{R}_{1}\oplus\mathcal{R}_{3/2}\oplus\mathcal{R}_{2}\,,
\end{empheq}
where $\mathcal{R}_{j}$ denotes a spin-$j$ representation of the $osp(1|2)$ superalgebra. In turn, these can be written as
\begin{empheq}{alignat=7}
	\mathcal{R}_{j}&=\mathcal{D}_{j-1/2}\oplus\mathcal{D}_{j}\,,
\end{empheq}
with $\mathcal{D}_{j}$ being a spin-$j$ representation of $sl(2)$. Thus, the bulk theory contains fields of spin\footnote{The bulk spin is $j+1$.} $(1,2,2,3)$ and $\left(\frac{3}{2},\frac{3}{2},\frac{5}{2},\frac{5}{2}\right)$, which under the holographic dictionary map to the scaling weights of the dual CFT operators. Of course, this coincides with the operator content $\left(J,T,V,W\right)$ and $\left(G^+,G^-,U^+,U^-\right)$ of the $\cW_{(3|2)}$ algebra.

Following the notation of appendix \ref{app: sl(3|2)}, we label the $sl(3|2)$ generators by $(J,L_i,A_i,W_m)$ and $(H_r,G_r,T_s,S_s)$. These correspond to $sl(2)$ multiplets of spin $(0,1,1,2)$ and $\left(\frac{1}{2},\frac{1}{2},\frac{3}{2},\frac{3}{2}\right)$, respectively. Then, according to the decomposition \eqref{Adjoint decomposition}, the Drinfeld-Sokolov connection \eqref{highest-weight} takes the explicit form
\begin{empheq}{alignat=7}\label{DS az}
	a_{z}&=L_{1}+\qL(z)L_{-1}+\qJ(z)J+\qA(z)A_{-1}+\qW(z)W_{-2}
	\cr
   	&+\qG(z)G_{-\frac{1}{2}}+\qH(z)H_{-\frac{1}{2}}+\qS(z)S_{-\frac{3}{2}}+\qT(z)T_{-\frac{3}{2}}\,.
\end{empheq}
Similarly, the gauge transformation parameter \eqref{gauge parameter general} becomes
\begin{empheq}{alignat=7}\label{gauge parameter}
	\Lambda&=\lJ(z)J + \lL(z) L_{1} + \lA(z) A_{1} + \lW(z) W_{2}
 	\cr
   	&+\lG(z)G_{\frac{1}{2}} +\lH(z)H_{\frac{1}{2}}+ \lS(z)S_{\frac{3}{2}}+\lT(z)T_{\frac{3}{2}}+\text{(16 higher-weight terms)}\,,
\end{empheq}
while the asymptotic symmetry condition \eqref{DS variation general} reads
\begin{empheq}{alignat=7}\label{DS variation}
	\partial\Lambda+[a_z,\Lambda]&=\delta\qL(z)L_{-1}+\delta\qJ(z)J+\delta\qA(z)A_{-1}+\delta\qW(z)W_{-2}
  	\cr
  	&+\delta\qG(z)G_{-\frac{1}{2}}+\delta\qH(z)H_{-\frac{1}{2}}+\delta\qS(z)S_{-\frac{3}{2}}+\delta\qT(z)T_{-\frac{3}{2}}\,.
\end{empheq}
It is important to emphasize that the currents $\qG(z)$, $\qH(z)$, $\qS(z)$ and $\qT(z)$, as well as the parameters $\lG(z)$, $\lH(z)$, $\lS(z)$ and $\lT(z)$ are Grassmann variables since they are associated with odd elements of the superalgebra.

Expression \eqref{DS variation} encodes $24$ equations, $16$ of which (the lower-weight components) determine the coefficients in front of the higher-weight generators in \eqref{gauge parameter}, with the remaining $8$ (the highest-weight components) allowing us to solve for the variations $\delta Q(z)$ in terms of the fields $Q(z)$, the parameters $\lambda(z)$ and their derivatives. For brevity, we omit the solution to the former.\footnote{The explicit form is needed to study the preserved symmetries of a given background. See \cite{Banados:2015tft} for an analysis of supersymmetric black holes.} We shall write the solution to the latter in a more convenient basis momentarily.
\subsubsection{Field redefinitions}\label{sec:fieldredef}
Were we to directly transform the field variations $\delta Q(z)$ obtained from \eqref{DS variation} into semiclassical OPE's we would find the symmetry algebra in a quite awkward form that obscures the superconformal structure discussed in section \ref{Section W32}. This is actually a rather generic feature of the asymptotic symmetry computations in Chern-Simons theory, present even in much simpler setups. One such example involves the bosonic theory based on the $sl(3)$ algebra with diagonally-embedded $sl(2)$, which results in the so-called $\cW_{3}^{(2)}$ algebra \cite{Ammon:2011nk,Campoleoni:2011hg}. In this case one finds that the naive bulk stress tensor (analogous to $\qL(z)$ in \eqref{DS az}) requires a Sugawara shift by the spin-1 current (analogous to $\qJ(z)$ in \eqref{DS az}) squared. A simultaneous redefinition of the naive infinitesimal $U(1)$ parameter (analogous to $\lJ$ in \eqref{gauge parameter}) is necessary in order for the $U(1)$ current to have the appropriate conformal dimension. In the present context we expect that even more involved modifications are needed because, in addition to a $U(1)$ current, there is a second bulk spin-2 field. Our goal in the reminder of this subsection is to provide the precise combinations of bulk fields $Q(z)$ and gauge transformation parameters $\lambda(z)$ such that the asymptotic symmetries of the $sl(3|2)$ Chern-Simons theory with AdS$_3$ boundary conditions take the form dictated by the semiclassical limit of the $\cN=2$ super-$\cW_3$ algebra discussed in section \ref{Section W32}. Not surprisingly, this turns out to be a laborious task, but there are a few guiding principles we can use to our advantage. 

The starting point to derive necessary field redefinitions is to recognize that the stress tensor and central charge in the dual CFT are given by
\begin{empheq}{alignat=7}\label{CFT stress tensor}
	T(z)&=-\frac{k_{cs}}{2}\text{sTr}\left[a_{z}^{2}\right]\,,
	&\qquad
	c&=12k_{cs}\text{sTr}\left[L_0^2\right]\,.
\end{empheq}
This bit of the holographic dictionary can be derived in multiple ways, e.g. by putting the Chern-Simons theory on a solid torus with modular parameter $\tau$ and studying the variation of the action under $\tau \to \tau +\delta \tau$ \cite{deBoer:2013gz,deBoer:2014fra}. Consequently, the redefinitions of fields should be such that the resulting combinations transform as primaries under this stress tensor. In the present case we find
\begin{empheq}{alignat=7}\label{trace stress tensor}
	T(z)&=\frac{c}{6}\left(\qL(z)+\frac{5}{3}\qA(z)+\qJ(z)^2\right)\,,
	&\qquad
	c&=18k_{cs}\,.
\end{empheq}
Here we see explicitly the Sugawara shift by the $U(1)$ current $\qJ(z)$. 

An additional clue comes from the observation that when bulk and CFT quantities are properly aligned, the connection \eqref{DS az} and gauge parameter \eqref{gauge parameter} should satisfy
\begin{equation}\label{trace connection gauge param}
	-k_{cs}\,\text{sTr}\left[a_{z}\Lambda\right]=\frac{c}{12}\partial^{2}\epsilon(z)+2\epsilon(z)T(z)+\sum_{s}s\,\epsilon_{s}(z)J_{s}(z)\,,
\end{equation}
where $\epsilon(z)$ parameterizes infinitesimal conformal transformations, $J_{s}(z)$ denotes a current of weight $s\,$, $\epsilon_{s}(z)$ is the associated infinitesimal parameter, and the sum runs over all the spins present in the spectrum (minus the stress tensor itself, which is singled out). This relation has been shown to be valid in the bosonic theory based on the $sl(N)$ algebra \cite{deBoer:2014fra}, even in non-principal embeddings where $U(1)$ currents are involved. We will verify that it remains true for $sl(3|2)$ as well. Using \eqref{trace stress tensor} we get
\begin{empheq}{alignat=7}
	-k_{cs}\,\text{sTr}\left[a_{z}\Lambda\right]&=\frac{c}{12}\left(\partial^2\lL(z)+\frac{5}{3}\partial^2\lA(z)\right)+2\left(\lL(z)+\frac{5}{3}\lA(z)\right)T(z)+\cdots\,,
\end{empheq}
from where we infer that the parameter of conformal transformations is
\begin{empheq}{alignat=7}
	\epsilon(z)&=\lL(z)+\frac{5}{3}\lA(z)\,.
\end{empheq}
One also learns from this calculation that the $U(1)$ parameter $\lJ(z)$ does in fact need to be modified as a result of the Sugawara shift in the stress tensor \eqref{trace stress tensor}, although we have not written it explicitly here since additional changes are required due to the presence of the other fields and parameters.

The final guiding principle in the construction of the holographic dictionary is the spectral flow automorphism \eqref{spectral flow currents W1} and \eqref{spectral flow currents W2} of the $\cW_{(3|2)}$ algebra. As expected, this piece of information is properly encoded in the symmetries of the Chern-Simons theory, being implemented by a gauge transformation associated with the $U(1)$ generator $J\in sl(3|2)$. Indeed, it is easy to see that a (finite) transformation with parameter $\Lambda(z)=\lJ(z)J$ induces the change\footnote{Notice that since $z\sim z+2\pi$, the gauge transformation with $\lJ(z)=i\eta z$ is singular except for integer or half-integer $\eta$.}
\begin{empheq}{alignat=7}\label{eq:Jtrans}
	\qJ(z)'&=\qJ(z)+\partial\lJ(z)\,,
\end{empheq}
which for $\lJ(z)\sim \eta z$ resembles the second equation in \eqref{spectral flow cylinder currents W1}. This naturally leads to the identification of $\qJ(z)$ with the CFT current $J(z)$. Moreover, since the $sl(2)$ multiplets $L_i$, $A_i$ and $W_m$ in $sl(3|2)$ commute with $J$, the corresponding fields in the Drinfeld-Sokolov connection \eqref{DS az} are inert under this transformation, that is, 
\begin{empheq}{alignat=7}
	\qL(z)'&=\qL(z)\,,
	&\qquad
	\qA(z)'&=\qA(z)\,,
	&\qquad
	\qW(z)'&=\qW(z)\,.
\end{empheq}
As a consequence, the map between these charges and their dual CFT variables can only involve spectral flow-invariant combinations such as \eqref{spectral flow invariant operators}. As for the fermions one finds that
\begin{equation}
\begin{aligned}
	\qG(z)'&=e^{-\lJ(z)}\qG(z)\,,
	&\qquad
	\qS(z)'&=e^{-\lJ(z)}\qS(z)\,,
	\\
	\qH(z)'&=e^{\lJ(z)}\qH(z)\,,
	&\qquad
	\qT(z)'&=e^{\lJ(z)}\qT(z)\,.
\end{aligned}
\end{equation}
Looking at \eqref{spectral flow cylinder currents W2}, spin and charge assignments then clearly imply that $\qH(z)\sim G^+(z)$, $\qG(z)\sim G^-(z)$, $\qT(z)\sim U^+(z)$ and $\qS(z)\sim U^-(z)$. This way, the spectral flow invariance of the dual CFT severely restricts the form that the bulk/boundary map can take.

Taking all of these insights into account, and after a detailed look at the transformation rules for the different charges, we are led to the following redefinitions of bulk fields:
\begin{equation}\label{dictionary currents}
\boxed{
\begin{aligned}
	\qJ(z)&=\frac{3}{c}J(z)\,,
	&\qquad
	\qH(z)&=-\frac{3}{c}G^{+}(z)\,,
	\\
	\qL(z)&=\frac{6}{c}\left(T(z)-\frac{3}{2c}J^{2}(z)+\frac{\kappa}{2}V(z)\right)\,,
	&\qquad
	\qG(z)&=\frac{3}{c}G^{-}(z)\,,
	\\
	\qA(z)&=-\frac{9\kappa}{5c}V(z)\,,
	&\qquad
	\qT(z)&=\frac{4\kappa}{5c}U^{+}(z)\,,
 	\\
	\qW(z)&=\frac{3\kappa}{5 c}\left(W(z) - \frac{6}{c}J(z)V(z)\right)\,,
	&\qquad
	\qS(z)&=-\frac{4\kappa}{5c}U^{-}(z)\,.
\end{aligned}
}
\end{equation}
Accordingly, the gauge parameters must be redefined as:
\begin{equation}\label{dictionary parameters}
\boxed{
\begin{aligned}
	\lJ(z)&=\lU(z)+\frac{3}{c}\epsilon(z)J(z)+\frac{6}{c}\chi(z)V(z)\,,
	&\qquad
	\lH(z)&=\lGm(z)\,,
	\\
	\lL(z)&=\epsilon(z)+\frac{\kappa}{2}\left(\lV(z)+\frac{6}{c}\chi(z)J(z)\right)\,,
	&\qquad
	\lG(z)&=-\lGp(z)\,,
	\\
	\lA(z)&=-\frac{3\kappa}{10}\left(\lV(z)+\frac{6}{c}\chi(z)J(z) \right)\,,
	&\qquad
	\lT(z)&=-\frac{2\kappa}{5}\lUm(z)\,,
	\\
	\lW(z)&=\frac{3\kappa}{10}\chi(z)\,,
	&\qquad
	\lS(z)&=\frac{2\kappa}{5}\lUp(z)\,.
\end{aligned}
}
\end{equation}
Condition \eqref{trace connection gauge param} is then satisfied only if the constant $\kappa$ appearing above is given by
\begin{empheq}{alignat=7}
	\kappa&=\pm\frac{5i}{2}\,.
\end{empheq}
Not coincidentally, this corresponds to the $c\to\infty$ limit of \eqref{definition kappa}. The transformation parameters $\epsilon_{s}(z)$ associated to the symmetries generated by each conserved current are identified as in table \ref{table current/parameter}. 
\begin{table}[h!]
\begin{center}
\begin{tabu}{c|[1.5pt]c}
\textrm{Current $J_s$}  & \textrm{Parameter $\epsilon_s$} \\ 
\tabucline[1.5pt]{-}\rule[-2ex]{0pt}{5ex} $J(z)$ & $\eta(z)$ \\ 
\hline 
\rule[-2ex]{0pt}{5ex} $T(z)$ & $\epsilon(z)$ \\ 
\hline 
\rule[-2ex]{0pt}{5ex} $G^{\pm}(z)$ & $\alpha^{\mp}(z)$ \\ 
\hline 
\rule[-2ex]{0pt}{5ex} $V(z)$ & $\gamma(z)$ \\ 
\hline 
\rule[-2ex]{0pt}{5ex} $U^{\pm}(z)$ & $\beta^{\mp}(z)$ \\ 
\hline 
\rule[-2ex]{0pt}{5ex} $W(z)$ & $\chi(z)$ \\ 
\end{tabu} 
\end{center}
\caption{Pairing between currents and  infinitesimal transformation parameters of the superconformal symmetry.}
\label{table current/parameter}
\end{table}

Relations \eqref{dictionary currents} and \eqref{dictionary parameters} are one of the main results of this paper. They constitute the first piece of the holographic dictionary, establishing the map between bulk currents and symmetry parameters in the $sl(3|2)$ Chern-Simons theory and their boundary CFT counterparts, and completing the partial analysis in \cite{Lu:1991ux,Tan:2012xi,Datta:2013qja}. Of course, the ultimate check of this result is the agreement between the OPE algebra for the redefined currents and the semiclassical limit of the $\cN=2$ super-$\cW_3$ algebra reviewed in section \ref{Section W32}.
\subsubsection{Variations and semiclassical OPE's}
In order to exhibit the transformation rules of the redefined fields under the asymptotic symmetries, it is convenient to introduce the operator
\begin{equation}\label{definition M}
M_{s,s'}^{s''}\left(\lambda; \phi\right) \equiv \sum_{i=1}^{s+s'-s''}\frac{\frac{(s+s'-s''-1)!}{(s+s'-s''-i)!}\frac{(s+s'+s''-2)!}{(s+s'+s''-i-1)!}}{\frac{(2s-2)!}{(2s-i-1)!}}\,\frac{\partial^{(i-1)}\lambda}{(i-1)!}\,\partial^{(s+s'-s''-i)}\phi\,.
\end{equation}
Its interpretation is as follows: $M_{s,s'}^{s"}(\lambda;\phi)$ gives the contribution of a field $\phi$ of spin-$s''$ to the variation of a spin-$s'$ primary under the symmetry generated by a spin-$s$ primary with associated infinitesimal parameter $\lambda$. The reason this particular operator simplifies the task of writing down the field variations is that the structure constants appearing in the OPE's of Virasoro primaries are constrained by the $sl(2,\mathds{R})$ covariance of the algebra; the coefficients appearing in \eqref{definition M} are then related to Clebsch-Gordan coefficients. 

Using the above notation, and in terms of the redefined charges and parameters \eqref{dictionary currents} and \eqref{dictionary parameters}, the asymptotic symmetry variations coming from \eqref{DS variation} are found to be
\begin{flalign*}
	\numberthis
	\label{var charges start}
	\delta J&=\frac{c}{3}\partial\lU+\epsilon\partial J+\partial\epsilon J+2\chi\partial V+2\partial\chi V+\lGm G^{-}-\lGp G^{+}+\lUm U^{-}-\lUp U^{+}\,,&
	\\
	\delta T&=\partial\lU J+\epsilon\partial T+2\partial\epsilon T+\frac{c}{12}\partial^{3}\epsilon+\lV\partial V+2\partial\lV V+2\chi\partial W+3\partial\chi W
	\\
	&+\frac{1}{2}\lGm\partial G^{-}+\frac{3}{2}\partial\lGm G^{-}+\frac{1}{2}\lGp\partial G^{+} +\frac{3}{2}\partial\lGp G^{+}
 	\\
 	\numberthis
	&+\frac{3}{2}\lUm\partial U^{-}+\frac{5}{2}\partial\lUm U^{-}+\frac{3}{2}\lUp\partial U^{+}+\frac{5}{2}\partial\lUp U^{+}\,,
    \\
   	\delta V&=\epsilon\partial V+2\partial\epsilon V+\frac{c}{12}\partial^{3}\lV+M_{2,2}^{2}\left(\lV;\cA{2}\right)
   	\\
   	&+6M_{3,2}^{1}\left(\chi;\cC{1}\right)+2M_{3,2}^{3}\left(\chi;\cC{3}\right)-M_{3,2}^{4}\left(\chi;\cC{4}\right)+\lGm U^{-}-\lGp U^{+}
   	\\
   	&-3M_{\frac{5}{2},2}^{\frac{3}{2}}\left(\lUm;\cPhi{3/2}{-}\right)+\frac{3}{2}M_{\frac{5}{2},2}^{\frac{5}{2}}\left(\lUm;\cPhi{5/2}{-}\right)-M_{\frac{5}{2},2}^{\frac{7}{2}}\left(\lUm;\cPhi{7/2}{-}\right)
   	\\\numberthis\label{var V}
   	&-3M_{\frac{5}{2},2}^{\frac{3}{2}}\left(\lUp;\cPhi{3/2}{+}\right)+\frac{3}{2}M_{\frac{5}{2},2}^{\frac{5}{2}}\left(\lUp;\cPhi{5/2}{+}\right)-M_{\frac{5}{2},2}^{\frac{7}{2}}\left(\lUp;\cPhi{7/2}{+}\right)\,,
  	\\
    \delta W&=2\partial\lU V+\epsilon\partial W+3\partial\epsilon W+6\partial^{3}\lV\cC{1}+M_{2,3}^{3}\left(\lV;\cC{3}\right)+\lV\cC{4}
    \\
    &+2M_{3,3}^{2}\left(\chi;\cB{2}\right)+M_{3,3}^{4}\left(\chi;\cB{4}\right)+\frac{c}{48}\partial^{5}\chi
    \\
    &+\frac{1}{2}M_{\frac{3}{2},3}^{\frac{5}{2}}\left(\lGm;U^{-}\right)+\frac{1}{2}M_{\frac{3}{2},3}^{\frac{5}{2}}\left(\lGp;U^{+}\right)+M_{\frac{5}{2},3}^{\frac{3}{2}}\left(\lUp;\cPsi{3/2}{+}\right)
    \\
    &+M_{\frac{5}{2},3}^{\frac{5}{2}}\left(\lUp;\cPsi{5/2}{+}\right)+\frac{3}{2}M_{\frac{5}{2},3}^{\frac{7}{2}}\left(\lUp;\cPsi{7/2}{+}\right)+\lUp \cPsi{9/2}{+}+M_{\frac{5}{2},3}^{\frac{3}{2}}\left(\lUm; \cPsi{3/2}{-}\right)
    \\
    \numberthis
    &+M_{\frac{5}{2},3}^{\frac{5}{2}}\left(\lUm;\cPsi{5/2}{-}\right)+\frac{3}{2}M_{\frac{5}{2},3}^{\frac{7}{2}}\left(\lUm;\cPsi{7/2}{-}\right)+\lUm \cPsi{9/2}{-}\,,
	\\
	\delta G^{\pm}&=\pm\lU G^{\pm}+\epsilon\partial G^{\pm}+\frac{3}{2}\partial\epsilon G^{\pm}\pm\lV U^{\pm}+2\chi\partial U^{\pm}+\frac{5}{2}\partial\chi U^{\pm}
   	\\
   	\numberthis
   	&+\lGmp\left(2T\mp\partial J\right)\mp2\partial\lGmp J+\frac{c}{3}\partial^{2}\lGmp+\lUmp\left(2W\mp3\partial V\right)\mp4\partial\lUmp V\,,
   	\\
   	\delta U^{\pm}&=\pm\lU U^{\pm}+\epsilon\partial U^{\pm}+\frac{5}{2}\partial\epsilon U^{\pm}+\lV\cPhi{7/2}{\pm}+M_{2,\frac{5}{2}}^{\frac{5}{2}}\left(\lV;\cPhi{5/2}{\pm}\right)+M_{2,\frac{5}{2}}^{\frac{3}{2}}\left(\lV;\cPhi{3/2}{\pm}\right)
   	\\
   	&-\chi\cPsi{9/2}{\pm}+2M_{3,\frac{5}{2}}^{\frac{7}{2}}\left(\chi;\cPsi{7/2}{\pm}\right)-2M_{3,\frac{5}{2}}^{\frac{5}{2}}\left(\chi;\cPsi{5/2}{\pm}\right)+4M_{3,\frac{5}{2}}^{\frac{3}{2}}\left(\chi;\cPsi{3/2}{\pm}\right)
   	\\
   	&+\lGmp\left(2W\mp\partial V\right)\mp4\partial\lGmp V+\frac{c}{12}\partial^{4}\lUmp+\lUmp\cD{4}+\lUpm\cE{4}{\pm}
    \\
    \numberthis
   	\label{var charges finish}
    &\mp M_{\frac{5}{2},\frac{5}{2}}^{1}\left(\lUmp;\cD{1}\right)+3 M_{\frac{5}{2},\frac{5}{2}}^{2}\left(\lUmp;\cD{2}\right)\mp M_{\frac{5}{2},\frac{5}{2}}^{3}\left(\lUmp;\cD{3}\right)\,,
\end{flalign*}
where the fields $\cA{s}$, $\cB{s}$, $\cC{s}$, $\cD{s}$, $\cE{s}{\pm}$, $\cPhi{s}{\pm}$, $\cPsi{s}{\pm}$ are given by
\begin{flalign*}
	\numberthis
	\label{semiclassical composites start}
	\cA{2}&=T-\frac{3}{2c}J^{2}+\kappa V\,,&
	\\
	\numberthis
	\cB{2}&=\frac{1}{20}\left(5T-\frac{3}{2c}J^2\right)+\frac{\kappa}{20}V\,,&
	\\
	\numberthis
	\cB{4}&=\frac{3}{2c}\Bigg\{16T^2+\frac{7}{2}\left(\partial G^+G^--G^+\partial G^-\right)+\frac{24}{c}\left(JG^+G^--J^2T\right)
	\\
	&+3\left(J\partial^2J-\frac{3}{10}\partial^2J^2\right)\Bigg\}+\frac{3\kappa}{5c}\Bigg\{16TV+\frac{11}{2}\left(G^-U^+-G^+U^-\right)+6JW\Bigg\}\,,
	\\
	\numberthis
	\cC{1}&=\textcolor{red}{\frac{1}{6}}\times\frac{1}{2}J\,,&
	\\
	\numberthis
	\cC{3}&=\frac{1}{2c}\Bigg\{\frac{15}{2}G^+G^-+8JT-\frac{12}{c}J^3\Bigg\}+\frac{\kappa}{5}\Bigg\{W+\frac{14}{c}JV\Bigg\}\,,
	\\
	\numberthis
	\cC{4}&=\frac{2}{c}\left(J\partial T-2\partial J T\right)+\frac{\kappa}{c}\Bigg\{2\left(J\partial V-2\partial JV\right)-3\left(G^+U^-+G^-U^+\right)\Bigg\}\,,
	\\
	\numberthis
	\cD{1}&=\frac{1}{4}J\,,&
	\\
	\numberthis
	\cD{2}&=\frac{1}{10}\left(5T-\frac{3}{c}J^2\right)+\frac{\kappa}{5}V\,,
	\\
	\numberthis
	\cD{3}&=\frac{3}{2c}\Bigg\{10JT-\frac{12}{c}J^3+\frac{13}{2}G^+G^-\Bigg\}+\frac{2\kappa}{5}\Bigg\{\frac{21}{c}JV-W\Bigg\}\,,
	\\
	\numberthis
	\cD{4}&=\frac{3}{c}\Bigg\{9T^2+\frac{5}{4}\left(\partial G^+G^--G^+\partial G^-\right)+\frac{12}{c}\left(JG^+G^--J^2T\right)
	\\
	&+\frac{1}{4}\left(J\partial^2J-\frac{3}{10}\partial^2J^2\right)\Bigg\}+\frac{12\kappa}{5c}\Bigg\{9TV+2\left(G^-U^+-G^+U^-\right)-JW\Bigg\}\,,
	\\
	\numberthis
	\cE{4}{\pm}&=-\frac{6}{c}\partial G^{\pm}G^{\pm}\mp\frac{12\kappa}{c}G^{\pm}U^{\pm}\,,&
	\\
	\numberthis
	\cPhi{3/2}{\pm}&=\pm\frac{1}{4}G^{\pm}\,,&
	\\
	\numberthis
	\cPhi{5/2}{\pm}&=-\frac{6}{5c}JG^{\pm}+\frac{2\kappa}{5}U^{\pm}\,,
	\\
	\numberthis
	\cPhi{7/2}{\pm}&=\textcolor{red}{2}\times\frac{3}{4c}\Bigg\{\pm9TG^{\pm}\mp\frac{12}{c}J^2G^{\pm}-\frac{1}{10}\left(2J\partial G^{\pm}-3\partial JG^{\pm}\right)\Bigg\}
	\\
	&\pm\frac{6\kappa}{5c}\Bigg\{9VG^{\pm}-JU^{\pm}\Bigg\}\,,
	\\
	\numberthis
	\cPsi{3/2}{\pm}&=\frac{1}{8}G^{\pm}\,,&
	\\
	\numberthis
	\cPsi{5/2}{\pm}&=\mp\frac{3}{10c}JG^{\pm}\pm\frac{\kappa}{10}U^{\pm}\,,
	\\
	\numberthis
	\cPsi{7/2}{\pm}&=\frac{3}{14c}\Bigg\{55TG^{\pm}-\frac{84}{c}J^2G^{\pm}\mp\frac{47}{10}\left(2J\partial G^{\pm}-3\partial JG^{\pm}\right)\Bigg\}
	\\
	&+\frac{6\kappa}{35c}\Bigg\{23VG^{\pm}+13JU^{\pm}\Bigg\}\,,
	\\
	\numberthis
	\label{semiclassical composites finish}
	\cPsi{9/2}{\pm}&=\frac{3}{7c}\Bigg\{2\left(3\partial TG^{\pm}-4T\partial G^{\pm}\right)-\left(\pm2\partial^2JG^{\pm}\mp4\partial J\partial G^{\pm}\pm J\partial^2G^{\pm}\right)\Bigg\}
	\\
	&+\frac{3\kappa}{7c}\Bigg\{\pm14TU^{\pm}\mp14WG^{\pm}+3\partial VG^{\pm}-4V\partial G^{\pm}+2J\partial U^{\pm}-5\partial JU^{\pm}\Bigg\}\,.
\end{flalign*}
It is reassuring to verify that expressions \eqref{semiclassical composites start}-\eqref{semiclassical composites finish}, which emerge entirely from a bulk analysis, correspond precisely to the semiclassical limit, taken as explained in section \ref{subsec: semiclassical}, of the full quantum composites \eqref{quantum composites start}-\eqref{quantum composites finish} that appear in the $\cW_{(3|2)}$ OPE algebra.\footnote{Compared to \cite{Romans:1991wi}, the modifications highlighted in red are necessary to match the holographic description.} At this point it also becomes clear that the fields defined through \eqref{dictionary currents} have conformal dimensions $\Delta$ and $U(1)$ charges $q$ as given in table \ref{new fields}, in agreement with the $\cN=2$ multiplet structure described in section \ref{Section W32}. These two facts are a non-trivial test for the validity of our results.
\begin{table}[h!]
\begin{center}
\begin{tabu}{c|[1.5pt]c|c}
\rule[-1ex]{0pt}{2.5ex} & $\Delta $ & $q$ \\ 
\tabucline[1.5pt]{-}\rule[-2ex]{0pt}{5ex} $J(z)$ & $1$ & $0 $ (anom) \\ 
\hline 
\rule[-2ex]{0pt}{5ex} $T(z)$ & $2$ (anom) & 0 \\ 
\hline 
\rule[-2ex]{0pt}{5ex} $G^{\pm}(z)$ & $3/2$ & $\pm 1$ \\ 
\hline 
\rule[-2ex]{0pt}{5ex} $V(z)$ & $2$ & $0$ \\ 
\hline 
\rule[-2ex]{0pt}{5ex} $U^{\pm}(z)$ & $5/2$ & $\pm 1$ \\ 
\hline 
\rule[-2ex]{0pt}{5ex} $W(z)$ & $3$ & $0$ \\ 
\end{tabu} 
\end{center}
\caption{Conformal dimensions and $U(1)$ charges of the redefined currents.}
\label{new fields}
\end{table}

The final step in identifying the asymptotic symmetry algebra is to convert the variations \eqref{var charges start}-\eqref{var charges finish} into semiclassical OPE's using Noether's theorem. To this purpose, following the assignments exhibited in table \ref{table current/parameter}, we define the total current
\begin{empheq}{alignat=7}
	J_{tot}(z)&=\lU(z)J(z)+\epsilon(z)T(z)+\lV(z)V(z)+\chi(z)W(z)
	 \cr
	&+\lGp(z)G^{+}(z)+\lGm(z)G^{-}(z)+\lUp(z)U^{+}(z)+\lUm(z)U^{-}(z)\,.
\end{empheq}
Then, after mapping the boundary cylinder to the complex plane via $z\to e^{iz}$, the transformations can be rewritten as
\begin{equation}
	\delta \mathcal{O}(w)=\oint_w\frac{dz}{2\pi i}J_{tot}(z)\mathcal{O}(w)\,,
\end{equation}
expression from which the OPE algebra can be read. For example, setting all the transformation parameters but $\lV$ to zero, the variation $\delta V$  in \eqref{var V} becomes
\begin{empheq}{alignat=7}
	\frac{c}{12}\partial^{3}\lV(w)+2\partial\lV(w)\cA{2}(w)+\lV(w)\partial\cA{2}(w)=\oint_w\frac{dz}{2\pi i}\lV(z)V(z)V(w)\,,
\end{empheq}
leading to the $VV$ OPE
\begin{empheq}{alignat=7}
	V(z)V(w)&\sim\frac{c/2}{(z-w)^{4}}+\left(\frac{2}{(z-w)^{2}}+\frac{1}{z-w}\partial\right)\cA{2}(w)\,.
\end{empheq}
We have checked that, quite satisfactorily, the full OPE algebra derived from the asymptotic symmetry variations \eqref{var charges start}-\eqref{var charges finish} is given precisely by the semiclassical limit of the $\cN=2$ super-$\cW_3$ algebra written in \eqref{Super Virasoro start}-\eqref{Super Virasoro finish}, \eqref{W1Ws OPE's start}-\eqref{W1Ws OPE's finish} and \eqref{W2W2 OPE's start}-\eqref{W2W2 OPE's finish}. This shows that the identifications \eqref{dictionary currents} and \eqref{dictionary parameters} between bulk and CFT quantities are indeed correct.
\subsection{Dictionary part II: sources and Ward identities}\label{subsec:sources}
Having successfully aligned bulk and boundary currents $J_s$ and infinitesimal parameters $\epsilon_s$,  we now turn to the study of sources (or chemical potentials). These play a central role in the discussion of black hole solutions and their thermodynamics \cite{Gutperle:2011kf,Chen:2013oxa,Datta:2013qja,Castro:2016tlm,Banados:2016hze,Kraus:2011ds,Gaberdiel:2012yb,Banados:2012ue,deBoer:2013gz,Ammon:2012wc,Bunster:2014mua,Perez:2013xi,Henneaux:2013dra,Perez:2012cf,Banados:2015tft}. We closely follow the work of \cite{deBoer:2014fra}, which offered a detailed account of sources in the context of the AdS$_{3}$/CFT$_{2}$ correspondence. As before, the treatment of the two connections $A$ and $\bar{A}$ is completely analogous, so we focus on the unbarred sector for concreteness.

Consider deforming a two-dimensional CFT with $\cW$-symmetry by coupling the set of (would-be) conserved higher spin currents $J_s(z,\bar{z})$ to some external fields $\mu_s(z,\bar{z})$. Restricting to chiral deformations, one natural possibility is to perturb the CFT action by
\begin{equation}\label{Action deformation}
	S=S_{CFT}+\int d^{2}z\,\sum_{s}\mu_{s}J_{s}\,.
\end{equation}
Another is to  write the deformed Hamiltonian
\begin{equation}\label{Hamiltonian deformation}
	H=H_{CFT}+\oint d\phi\,\sum_{s}\mu_{s}J_{s}\,.
\end{equation}
In either case, the $\mathcal{W}$-symmetry is still realized at the level of the partition function provided that one transforms the sources $\mu_s(z,\bar{z})$ accordingly \cite{deBoer:2014fra}. This results in the existence of Ward identities for the one-point functions of the currents in the presence of sources.

It is common knowledge that a holographic description of these deformations requires generalizing the AdS boundary conditions \eqref{highest-weight}, such that the bulk Chern-Simons fields now include the deformation parameters in their asymptotics. As argued in \cite{deBoer:2014fra}, the ensuing structure is best described in terms of a ``Drinfeld-Sokolov pair'', consisting of one component of the connection $a(z,\bar{z})$ carrying the bulk currents as highest-weights, and a conjugate component carrying the corresponding sources as lowest-weights. Since all source-free solutions satisfy $a_{\bar{z}}=0$, sources should certainly be included in this component of the connection. However, in their presence, the question arises of whether the currents should be incorporated in $a_{z}$ or in $a_z+a_{\bar{z}}\neq a_z$. This leads to two natural choices of boundary conditions for the gauge fields, namely, \emph{holomorphic} boundary conditions, given by the Drinfeld-Sokolov pair
\begin{empheq}{alignat=7}
	\label{holomorphic bc}
	a_z&=L_1+\bm Q(z,\bar{z})\,,
	&\qquad
	2a_{\bar{z}}&=\bm\nu(z,\bar{z})+\cdots\,,
\end{empheq}
and \emph{canonical} boundary conditions, implemented by
\begin{empheq}{alignat=7}
	\label{canonical bc}
	a_z+a_{\bar{z}}&=L_1+\bm Q(z,\bar{z})\,,
	&\qquad
	2a_{\bar{z}}&=\bm\nu(z,\bar{z})+\cdots\,.
\end{empheq}
The matrices $\bm Q$ and $\bm\nu$ are such that $[L_{-1},\bm Q]=0$ and $[L_1,\bm\nu]=0$ (highest and lowest-weight, respectively), and the dots represent higher-weight terms that are fixed algebraically by the flatness condition \eqref{flatness}. Of course, the source-free solution \eqref{highest-weight} is recovered for $\bm\nu=0$. 

From the field theory point of view, it was shown in \cite{Gutperle:2011kf,Ammon:2011nk,deBoer:2014fra} that holomorphic boundary conditions correspond precisely to deformations \eqref{Action deformation} of the CFT action, whereas the the canonical choice maps to deformations \eqref{Hamiltonian deformation} of the Hamiltonian. In either case the connection $a(z,\bar{z})$ is no longer holomorphic, and the bulk equations of motion become the CFT's Ward identities. Below we will exemplify in detail the incorporation of sources in the holomorphic case, where they are more symmetrical. Then we will point out the changes needed to accomplish this in the canonical case.
\subsubsection{Action deformations and holomorphic boundary conditions}
Most of the analysis of asymptotic symmetries in the previous section consisted in finding the correct combinations of bulk currents in $\bm Q$ such that the transformation rules took the form dictated by the $\cW_{(3|2)}$ algebra. By the same token, the analysis of deformed boundary conditions boils down to finding the precise combination of bulk sources in $\bm\nu$ such that the equations of motion for the Drinfeld-Sokolov pair of connections reproduce the CFT's Ward identities. Happily, in the case of holomorphic boundary conditions \eqref{holomorphic bc}, all of the necessary algebra can be recycled from the asymptotic symmetry calculations by noticing that the flatness equation \eqref{flatness}, written as
\begin{empheq}{alignat=7}
	\partial a_{\bar{z}}+[a_z,a_{\bar{z}}]&=\bar{\partial}a_z\,,
\end{empheq}
is essentially the same as equation \eqref{DS variation general} for the variations $\delta\bm Q$ and transformation parameters $\Lambda$, with the replacements $\delta\bm Q\to\bar{\partial}\bm Q$ and $\Lambda\to a_{\bar{z}}$. Moreover, the condition
\begin{empheq}{alignat=7}\label{trace holomorphic DS pair}
	-k_{cs}\,\text{sTr}\bigl[a_za_{\bar{z}}\bigr]&=\frac{c}{12}\partial^{2}\mT(z,\bar{z})+2\mT(z,\bar{z})T(z,\bar{z})+\sum_{s}s\mu_{s}(z,\bar{z})J_{s}(z,\bar{z})\,,
\end{empheq}
which must also be satisfied when bulk and boundary quantities are properly aligned \cite{deBoer:2014fra}, follows from \eqref{trace connection gauge param} by the same replacements.

These considerations allow us to directly state the second piece of the holographic dictionary, namely, the one relating the chemical potentials in the $sl(3|2)$ Chern-Simons theory with the sources coupling to the conserved currents in a CFT displaying $\cW_{(3|2)}$ symmetry. Writing a lowest-weight ansatz similar to \eqref{gauge parameter},
\begin{empheq}{alignat=7}\label{azbar}
	a_{\bar{z}}&=\nJ(z,\bar{z})J+\nL(z,\bar{z})L_1+\nA(z,\bar{z})A_1+\nW(z,\bar{z})W_2
	\cr
	&+\nG(z,\bar{z})G_{\frac{1}{2}}+\nH(z,\bar{z})H_{\frac{1}{2}}+\nS(z,\bar{z})S_{\frac{3}{2}}+\nT(z,\bar{z})T_{\frac{3}{2}}+\textrm{(higher-weight terms)}\,,
\end{empheq}
we find that the correct combination of bulk fields coupling to the CFT currents is\footnote{We omit the dependence in $(z,\bar{z})$ for simplicity} (cf. \eqref{dictionary parameters})
\begin{equation}\label{dictionary sources}
\boxed{
\begin{aligned}
	\nJ&=\mJ+\frac{3}{c}\mT J+\frac{6}{c}\mW V\,,
	&\qquad
	\nH&=\mGm\,,
	\\
	\nL&=\mT+\frac{\kappa}{2}\left(\mV+\frac{6}{c}\mW J\right)\,,
	&\qquad
	\nG&=-\mGp\,,
	\\
	\nA&=-\frac{3\kappa}{10}\left(\mV+\frac{6}{c}\mW J\right)\,,
	&\qquad
	\nT&=-\frac{2\kappa}{5}\mUm\,,
	\\
	\nW&=\frac{3\kappa}{10}\mW\,,
	&\qquad
	\nS&=\frac{2\kappa}{5}\mUp\,,
\end{aligned}
}
\end{equation}
where $\kappa=\pm5i/2$ as before. Expression \eqref{DS az} for $a_z$, as well as the map \eqref{dictionary currents} between bulk and boundary charges, still apply in the deformed theory, albeit with an additional anti-holomorphic dependence. Importantly, with these redefinitions, the Drinfeld-Sokolov connection automatically verifies relation \eqref{trace holomorphic DS pair} for the principally-embedded $osp(1|2)\subset sl(3|2)$ spectrum, leading to the pairing of the different CFT sources and currents shown in table \ref{table current/source}.
\begin{table}[h!]
\begin{center}
\begin{tabu}{c|[1.5pt]c}
\textrm{Current} & \textrm{Source} \\ 
\tabucline[1.5pt]{-}\rule[-2ex]{0pt}{5ex} $J(z,\bar{z})$ & $\mJ(z,\bar{z})$ \\ 
\hline 
\rule[-2ex]{0pt}{5ex} $T(z,\bar{z})$ & $\mT(z,\bar{z})$ \\ 
\hline 
\rule[-2ex]{0pt}{5ex} $G^{\pm}(z,\bar{z})$ & $\mGmp(z,\bar{z})$ \\ 
\hline 
\rule[-2ex]{0pt}{5ex} $V(z,\bar{z})$ & $\mV(z,\bar{z})$ \\ 
\hline 
\rule[-2ex]{0pt}{5ex} $U^{\pm}(z,\bar{z})$ & $\mUmp(z,\bar{z})$ \\ 
\hline 
\rule[-2ex]{0pt}{5ex} $W(z,\bar{z})$ & $\mW(z,\bar{z})$ \\ 
\end{tabu} 
\end{center}
\caption{Correspondence between redefined currents and sources.}
\label{table current/source}
\end{table}

Finally, the solution to the highest-weight components of the equation of motion, that is, the solution for $\bar{\partial}\bm Q(z,\bar{z})$, follows directly from the transformation rules \eqref{var charges start}-\eqref{var charges finish}, yielding
\begin{flalign*}
	\bar{\partial}J&=\frac{c}{3}\partial\mJ+\mT\partial J+\partial\mT J+2\mW\partial V+2\partial\mW V
	\\
	\numberthis
	\label{holo Ward start}
	&+\mGm G^{-}-\mGp G^{+}+\mUm U^{-}-\mUp U^{+}\,,&
	\\
	\bar{\partial}T&=\partial\mJ J+\mT\partial T+2\partial\mT T+\frac{c}{12}\partial^{3}\mT+\mV\partial V+2\partial\mV V+2\mW\partial W+3\partial\mW W
	\\
	&+\frac{1}{2}\mGm\partial G^{-}+\frac{3}{2}\partial\mGm G^{-}+\frac{1}{2}\mGp\partial G^{+} +\frac{3}{2}\partial\mGp G^{+}
 	\\
 	\numberthis
	&+\frac{3}{2}\mUm\partial U^{-}+\frac{5}{2}\partial\mUm U^{-}+\frac{3}{2}\mUp\partial U^{+}+\frac{5}{2}\partial\mUp U^{+}\,,
    \\
   	\bar{\partial}V&=\mT\partial V+2\partial\mT V+\frac{c}{12}\partial^{3}\mV+M_{2,2}^{2}\left(\mV;\cA{2}\right)
   	\\
   	&+6M_{3,2}^{1}\left(\mW;\cC{1}\right)+2M_{3,2}^{3}\left(\mW;\cC{3}\right)-M_{3,2}^{4}\left(\mW;\cC{4}\right)+\mGm U^{-}-\mGp U^{+}
   	\\
   	&-3M_{\frac{5}{2},2}^{\frac{3}{2}}\left(\mUm;\cPhi{3/2}{-}\right)+\frac{3}{2}M_{\frac{5}{2},2}^{\frac{5}{2}}\left(\mUm;\cPhi{5/2}{-}\right)-M_{\frac{5}{2},2}^{\frac{7}{2}}\left(\mUm;\cPhi{7/2}{-}\right)
   	\\
   	 \numberthis
   	&-3M_{\frac{5}{2},2}^{\frac{3}{2}}\left(\mUp;\cPhi{3/2}{+}\right)+\frac{3}{2}M_{\frac{5}{2},2}^{\frac{5}{2}}\left(\mUp;\cPhi{5/2}{+}\right)-M_{\frac{5}{2},2}^{\frac{7}{2}}\left(\mUp;\cPhi{7/2}{+}\right)\,,
  	\\
    \bar{\partial}W&=2\partial\mJ V+\mT\partial W+3\partial\mT W+6\partial^{3}\mV\cC{1}+M_{2,3}^{3}\left(\mV;\cC{3}\right)+\mV\cC{4}
    \\
    &+2M_{3,3}^{2}\left(\mW;\cB{2}\right)+M_{3,3}^{4}\left(\mW;\cB{4}\right)+\frac{c}{48}\partial^{5}\mW
    \\
    &+\frac{1}{2}M_{\frac{3}{2},3}^{\frac{5}{2}}\left(\mGm;U^{-}\right)+\frac{1}{2}M_{\frac{3}{2},3}^{\frac{5}{2}}\left(\mGp;U^{+}\right)+M_{\frac{5}{2},3}^{\frac{3}{2}}\left(\mUp;\cPsi{3/2}{+}\right)
    \\
    &+M_{\frac{5}{2},3}^{\frac{5}{2}}\left(\mUp;\cPsi{5/2}{+}\right)+\frac{3}{2}M_{\frac{5}{2},3}^{\frac{7}{2}}\left(\mUp;\cPsi{7/2}{+}\right)+\mUp \cPsi{9/2}{+}+M_{\frac{5}{2},3}^{\frac{3}{2}}\left(\mUm; \cPsi{3/2}{-}\right)
    \\
    \numberthis
    &+M_{\frac{5}{2},3}^{\frac{5}{2}}\left(\mUm;\cPsi{5/2}{-}\right)+\frac{3}{2}M_{\frac{5}{2},3}^{\frac{7}{2}}\left(\mUm;\cPsi{7/2}{-}\right)+\mUm \cPsi{9/2}{-}\,,
	\\
	\bar{\partial}G^{\pm}&=\pm\mJ G^{\pm}+\mT\partial G^{\pm}+\frac{3}{2}\partial\mT G^{\pm}\pm\mV U^{\pm}+2\mW\partial U^{\pm}+\frac{5}{2}\partial\mW U^{\pm}
   	\\
   	\numberthis
   	&+\mGmp\left(2T\mp\partial J\right)\mp2\partial\mGmp J+\frac{c}{3}\partial^{2}\mGmp+\mUmp\left(2W\mp3\partial V\right)\mp4\partial\mUmp V\,,
   	\\
   	\bar{\partial}U^{\pm}&=\pm\mJ U^{\pm}+\mT\partial U^{\pm}+\frac{5}{2}\partial\mT U^{\pm}+\mV\cPhi{7/2}{\pm}+M_{2,\frac{5}{2}}^{\frac{5}{2}}\left(\mV;\cPhi{5/2}{\pm}\right)+M_{2,\frac{5}{2}}^{\frac{3}{2}}\left(\mV;\cPhi{3/2}{\pm}\right)
   	\\
   	&-\mW\cPsi{9/2}{\pm}+2M_{3,\frac{5}{2}}^{\frac{7}{2}}\left(\mW;\cPsi{7/2}{\pm}\right)-2M_{3,\frac{5}{2}}^{\frac{5}{2}}\left(\mW;\cPsi{5/2}{\pm}\right)+4M_{3,\frac{5}{2}}^{\frac{3}{2}}\left(\mW;\cPsi{3/2}{\pm}\right)
   	\\
   	&+\mGmp\left(2W\mp\partial V\right)\mp4\partial\mGmp V+\frac{c}{12}\partial^{4}\mUmp+\mUmp\cD{4}+\mUpm\cE{4}{\pm}
    \\\numberthis
    \label{holo Ward finish}
    &\mp M_{\frac{5}{2},\frac{5}{2}}^{1}\left(\mUmp;\cD{1}\right)+3 M_{\frac{5}{2},\frac{5}{2}}^{2}\left(\mUmp;\cD{2}\right)\mp M_{\frac{5}{2},\frac{5}{2}}^{3}\left(\mUmp;\cD{3}\right)\,.
\end{flalign*}
The composite fields $\cA{s}$, $\cB{s}$, $\cC{s}$, $\cD{s}$, $\cE{s}{\pm}$, $\cPhi{s}{\pm}$, $\cPsi{s}{\pm}$ and the operators $M_{s,s'}^{s"}(\lambda;\phi)$ are the same as in section \ref{subsec: asymp symm}. These are the holographic Ward identities corresponding to the deformation \eqref{Action deformation} of the CFT action.
\subsubsection{Hamiltonian deformations and canonical boundary conditions}\label{subsec:canonical bcs}
For the implementation of canonical boundary conditions it is convenient to switch to the real coordinates $\phi=\left(z+\bar{z}\right)/2$ and $t_E=i\left(z-\bar{z}\right)/2$, in terms of which the Drinfeld-Sokolov pair \eqref{canonical bc} becomes
\begin{empheq}{alignat=7}
	\label{canonical bc phi t}
	a_{\phi}&=L_1+\bm Q(t_E,\phi)\,,
	&\qquad
	a_{\phi}-ia_{t_E}&=\bm\nu(t_E,\phi)+\cdots\,.
\end{empheq}
Bulk charges and sources are then included as in \eqref{DS az} and \eqref{azbar}, but with $a_z\to a_{\phi}$ and $a_{\bar{z}}\to a_{\phi}-ia_{t_E}$. Since the holographic dictionary for the charges is derived from the source-free solutions, where $a_{\phi}=a_z$, we still have \eqref{dictionary currents}. Moreover, the Drinfeld-Sokolov pair now satisfies the relation \cite{deBoer:2014fra}
\begin{empheq}{alignat=7}
	-k_{cs}\,\text{sTr}\bigl[a_{\phi}\left(a_{\phi}-ia_{t_E}\right)\bigr]&=\frac{c}{12}\partial^{2}\mT(t_E,\phi)+2\mT(t_E,\phi)T(t_E,\phi)
	\cr
	&+\sum_{s}s\mu_{s}(t_E,\phi)J_{s}(t_E,\phi)\,,
\end{empheq}
leading to the same map \eqref{dictionary sources} for the sources, with the pairing displayed in table \ref{table current/source}. As explained in \cite{deBoer:2014fra}, the Ward identities corresponding to the Hamiltonian deformation \eqref{Hamiltonian deformation} follow from \eqref{holo Ward start}-\eqref{holo Ward finish} by replacing $\bar{\partial}\to i\partial_{t_E}-\partial_{\phi}$ and $\partial\to\partial_{\phi}$. 
For example, the canonical stress tensor Ward identity reads
\begin{align}
	i\partial_{t_E}T-\partial_{\phi}T&=\frac{c}{12}\partial_{\phi}^{3}\mT+\partial_{\phi}\mJ  J+\mT\partial_{\phi}T+2\partial_{\phi}\mT T
	\cr
	&+\mV\partial_{\phi}V+2\partial_{\phi}\mV V+2\mW\partial_{\phi}W+3\partial_{\phi}\mW W
	\cr
	&+\frac{1}{2}\mGm\partial_{\phi} G^-+\frac{3}{2}\partial_{\phi}\mGm G^-+\frac{1}{2}\mGp\partial_{\phi} G^++\frac{3}{2}\partial_{\phi}\mGp G^+
	\cr
	&+\frac{3}{2}\mUm\partial_{\phi}U^-+\frac{5}{2}\partial_{\phi} \mUm U^-+\frac{3}{2}\mUp\partial_{\phi}U^++\frac{5}{2}\partial_{\phi}\mUp U^+\,,
\end{align}
The remaining Ward identities are derived in a similar fashion.

We close this section by commenting that when putting the theory at finite temperature, i.e. on a torus, one also needs to specify how the thermal sources scale with the temperature; only then is the partition function (and consequently the free energy) well-defined \cite{deBoer:2014fra}. Moreover, in this context it is redundant to include the source for the stress tensor in the connection $a_{\phi}-ia_{t_E}$, since it can be incorporated as the modular parameter of the torus. From \eqref{dictionary sources}, we see that setting $\mu_2=0$ amounts to fixing $\nA=-\frac{3}{5}\nL$, so that only the source for the combination of generators $L_1-\frac{3}{5}A_1$ is turned on. These issues are discussed in our companion paper \cite{Banados:2015tft}, where the thermodynamics of black hole solutions in the $sl(3|2)$ theory with canonical boundary conditions were studied.
\subsection{Lorentzian connections and $su(2,1|1,1)$}\label{sec: Lorentzian}
The preceding construction of the holographic dictionary was carried out in the Euclidean formalism, where the Chern-Simons connection $a(z,\bar{z})$ is valued in the complex superalgebra $sl(3|2;\mathds{C})$. When continuing back to Lorentzian signature the question arises of which real form appropriately describes the CFT structure that we have uncovered. According to \cite{Frappat:1996pb}, the candidate superalgebras are
\begin{empheq}{alignat=5}
	&sl(3|2;\mathds{R})\supset sl(3;\mathds{R})\oplus sl(2;\mathds{R})\oplus\mathds{R}\,,
	\cr
	&sl(3|2;\mathds{H})\supset su^*(3)\oplus su^*(2)\oplus\mathds{R}\,,
	\cr
	&su(p,3-p|q,2-q)\supset su(p,3-p)\oplus su(q,2-q)\oplus i\mathds{R}\,.
\end{empheq}
As we will now see, the real form $su(2,1|1,1)$ is singled out after imposing the standard hermiticity conditions on the generators of the $\mathcal{N}=2$ super-$\cW_3$ algebra and applying the holographic map. To make the analysis more transparent we will focus on source-free solutions \eqref{highest-weight}; having understood the reality properties of the charges those of the sources follow directly.

All throughout section \ref{Section Dictionary} we have complied with the conventions of \cite{deBoer:2013gz,deBoer:2014fra}, which implemented the continuation of the Lorentzian higher spin theory to Euclidean signature by taking $t+\phi\to z$ and $t-\phi\to-\bar{z}$. In particular, the definition \eqref{CFT stress tensor} of the stress tensor and the condition \eqref{trace connection gauge param} between bulk and boundary quantities, both of which where crucial in the derivation of the holographic dictionary, pend on this choice. This means that our conventions are such that the $\cN=2$ super-$\cW_3$ generators on the cylinder satisfy
\begin{equation}\label{conjugation cylinder W32}
\begin{aligned}
	J(z)^{\dagger}&=-J\left(\bar{z}\right)\,,
	&\qquad
	G^+(z)^{\dagger}&=e^{\frac{i\pi}{2}}G^-\left(\bar{z}\right)\,,
	\\
	T(z)^{\dagger}&=T\left(\bar{z}\right)\,,
	&\qquad
	G^-(z)^{\dagger}&=e^{\frac{i\pi}{2}}G^+\left(\bar{z}\right)\,,
	\\
	\bar{\kappa}V(z)^{\dagger}&=\kappa V\left(\bar{z}\right)\,,
	&\qquad
	\bar{\kappa}U^+(z)^{\dagger}&=e^{-\frac{i\pi}{2}}\kappa U^-\left(\bar{z}\right)\,,
	\\
	\bar{\kappa}W(z)^{\dagger}&=-\kappa W\left(\bar{z}\right)\,,
	&\qquad
	\bar{\kappa}U^+(z)^{\dagger}&=e^{-\frac{i\pi}{2}}\kappa U^-\left(\bar{z}\right)\,,
\end{aligned}
\end{equation}
as follows from \eqref{conjugation cylinder} and \eqref{conjugation cylinder charged} with $\zeta=i$. Thus, in light of the map \eqref{dictionary currents}, the different components in the Drinfeld-Sokolov connection must obey the following reality conditions:
\begin{equation}\label{reality cylinder charges}
\begin{aligned}
	\overline{\qJ(z)}&=-\qJ(\bar{z})\,,
	&\qquad
	\overline{\qG(z)}&=e^{-\frac{i\pi}{2}}\qH(\bar{z})\,,
	\\
	\overline{\qL(z)}&=\qL(\bar{z})\,,
	&\qquad
	\overline{\qH(z)}&=e^{-\frac{i\pi}{2}}\qG(\bar{z})\,,
	\\
	\overline{\qA(z)}&=\qA(\bar{z})\,,
	&\qquad
	\overline{\qT(z)}&=e^{\frac{i\pi}{2}}\qS(\bar{z})\,,
	\\
	\overline{\qW(z)}&=-\qW(\bar{z})\,,
	&\qquad
	\overline{\qT(z)}&=e^{\frac{i\pi}{2}}\qS(\bar{z})\,.
\end{aligned}
\end{equation}

In order to figure out the appropriate real form of $sl(3|2)$ that is compatible with \eqref{reality cylinder charges} we seek for combinations of currents that are real when continued back to Lorentzian signature via $z\to x^+$. Then, the correct superalgebra will be the one spanned by those generators accompanying these charges in $a_z\to a_+$. With this in mind, a more suggestive way of writing the Drinfeld-Sokolov connection \eqref{DS az} is
\begin{empheq}{alignat=7}
	a_z&=L_1+iQ_1(z)J+Q_2(z)L_{-1}+\tilde{Q}_2(z)A_{-1}+iQ_3(z)W_{-2}
	\cr
	&+e^{\frac{i\pi}{4}}Q_{\frac{1}{2}}^+(z)\left(H_{-\frac{1}{2}}+G_{-\frac{1}{2}}\right)+e^{-\frac{i\pi}{4}}Q_{\frac{1}{2}}^-(z)\left(H_{-\frac{1}{2}}-G_{-\frac{1}{2}}\right)
	\cr
	&+e^{-\frac{i\pi}{4}}Q_{\frac{3}{2}}^+(z)\left(T_{-\frac{3}{2}}+S_{-\frac{3}{2}}\right)+e^{\frac{i\pi}{4}}Q_{\frac{3}{2}}^-(z)\left(T_{-\frac{3}{2}}-S_{-\frac{3}{2}}\right)\,,
\end{empheq}
where
\begin{equation}
\begin{aligned}
	Q_1(z)&=-i\qJ(z)\,,
	&\qquad
	Q_{\frac{1}{2}}^+(z)&=\frac{1}{2}e^{-\frac{i\pi}{4}}\left(\qH(z)+\qG(z)\right)\,,
	\\
	Q_2(z)&=\qL(z)\,,
	&\qquad
	Q_{\frac{1}{2}}^-(z)&=\frac{1}{2}e^{\frac{i\pi}{4}}\left(\qH(z)-\qG(z)\right)\,,
	\\
	\tilde{Q}_2(z)&=\qA(z)\,,
	&\qquad
	Q_{\frac{3}{2}}^+(z)&=\frac{1}{2}e^{\frac{i\pi}{4}}\left(\qT(z)+\qS(z)\right)\,,
	\\
	Q_3(z)&=-i\qW(z)\,,
	&\qquad
	Q_{\frac{3}{2}}^-(z)&=\frac{1}{2}e^{-\frac{i\pi}{4}}\left(\qT(z)-\qS(z)\right)\,.
\end{aligned}
\end{equation}
It is easy to check using \eqref{reality cylinder charges} that all these combinations of charges satisfy $\overline{Q(z)}=Q(\bar{z})$, implying that they are real if restricted to real arguments. In other words, $\overline{Q(x^+)}=Q(x^+)$. This way we learn that the CFT structure of the Euclidean theory dictates that a general Lorentzian connection on the cylinder takes values in the real form of the $sl(3|2)$ algebra generated by
\begin{equation}
\begin{array}{c}
	iJ\,,
	\qquad
	L_i\,,
	\qquad
	A_i\,,
	\qquad
	iW_m\,,
	\\
	{\displaystyle e^{\frac{i\pi}{4}}\left(H_r+G_r\right)}\,,	
	\qquad
	{\displaystyle e^{-\frac{i\pi}{4}}\left(H_r-G_r\right)}\,,
	\qquad
	{\displaystyle e^{-\frac{i\pi}{4}}\left(T_s+S_s\right)}\,,
	\qquad
	{\displaystyle e^{\frac{i\pi}{4}}\left(T_s-S_s\right)}\,,
\end{array}
\end{equation}
As discussed in appendix \ref{app: sl(3|2)}, this corresponds to the superalgebra $su(2,1|1,1)$.\footnote{Another way to derive this result, purely from the bulk perspective, is to start from  $a_z=L_1+\bm Q(z)$ and make the change of coordinates $z'=\zeta z$. Then, an additional gauge transformation with parameter $\Lambda=\zeta^{L_0}$ is necessary in order to bring the new connection $a_{z'}=\zeta^{-1}a_z$ back to Drinfeld-Sokolov form, $a'_{z'}=L_1+\bm Q'(z')$. The new charges read $\bm Q'(z')=\zeta^{-1}\zeta^{-L_0}\bm Q(z)\zeta^{L_0}$. Setting $\zeta=e^{\frac{i\pi}{2}}$, it follows that $\bm Q'\in sl(3|2;\mathds{R})\Leftrightarrow \bm Q\in su(2,1|1,1)$.}
\section{Conclusion}\label{Section Conclusion}
The purpose of this note was to provide the explicit form of the holographic dictionary between $sl(3|2)$ Chern-Simons supergravity on AdS$_3$ and two-dimensional CFT's with $\mathcal{N}=2$ super-$\mathcal{W}_3$ symmetry that was used in \cite{Banados:2015tft}. The main entries of the dictionary are the identification of bulk and boundary currents given in \eqref{dictionary currents}, with their corresponding transformation parameters \eqref{dictionary parameters}, and the identification of the Chern-Simons and CFT sources in \eqref{dictionary sources}. This allowed us to display in full detail the holographic Ward identities \eqref{holo Ward start}-\eqref{holo Ward finish}. The other entry is the identification of $su(2,1|1,1)$ as the correct real form of $sl(3|2)$ in Lorentzian signature. Along the way, we also corrected some typos in the original reference \cite{Romans:1991wi} on the $\cW_{(3|2)}$ algebra. These corrections were verified by three independent methods: i) fulfillment of the Jacobi identities for the OPE's, ii) spectral flow invariance of the OPE algebra, and iii) agreement with the asymptotic symmetries of Chern-Simons theory.

It is worth highlighting the role of the spectral flow automorphism in building the holographic dictionary. Our discussion in section \ref{sec:fieldredef} applies broadly to any gauge algebra $\mathfrak{g}$ that contains a $U(1)$ generator that would lead to spectral flow symmetry in the boundary $\cW$-algebra. In particular,  following the argument around \eqref{eq:Jtrans}, it should be straightforward to identify the components of $\bm Q(z)$ that are spectral flow invariant in a highest-weight gauge \eqref{highest-weight} for the connection. This, combined with \eqref{trace connection gauge param}, leads to a clear and simple basis in which to setup the dictionary with CFT variables. 

Naturally, some of our results can be extended to the analysis of Lorentzian solutions in the $sl(N|N-1)$ theory, where the same reasoning shows that the correct real form of the bulk gauge algebra that is consistent with the structure of $\cN=2$ supersymmetric higher spin symmetries is $su(p,N-p|q,N-1-q)\,$ and not $sl(N|N-1;\mathds{R})$ as naively expected. Perhaps the easiest way to see this is to note that $su(p,m-p|q,n-q)$ is the only real form of $sl(m|n;\mathds{C})$ that possesses a compact Abelian generator in the bosonic subalgebra \cite{Frappat:1996pb}. As shown \cite{Banados:2015tft}, this property is crucial for compatibility with $R$-charge quantization and the existence of Killing spinors with non-trivial angular dependence. Ultimately, this is tied to the fact that we have performed a Drinfeld-Sokolov reduction based on an $sl(2|1)$ embedding instead of an $osp(2|2)$ embedding as required to reproduce the $\cN=2$ CFT structure (see e.g. \cite{Tan:2012xi,Peng:2012ae}). To our knowledge, the issue of identifying the appropriate real form has not been discussed in the literature so far. It is, however, an important ingredient in the holographic dictionary if one is to match bulk and boundary results correctly.

An important asset of the dualities addressed in this paper is that they exploit the topological formulation of the Chern-Simons theory in order to set up and perform tractable calculations that are quite challenging in CFT's with $\cW$-algebra using solely field-theoretical techniques. The analysis here is one very modest example in the context of supersymmetric dualities. In recent years this asset has been applied to Wilson lines in Chern-Simons theory as one efficient approach to evaluate $\cW$-conformal blocks in the CFT \cite{deBoer:2014sna,Melnikov:2016eun,Hulik:2018dpl}. In our case it would be interesting to evaluate a supersymmetric Wilson line and include quantum corrections following the approach of \cite{Fitzpatrick:2016mtp,Besken:2017fsj,Hikida:2017ehf,Hikida:2018dxe,Hikida:2018eih,Besken:2018zro}. This would allow us to  improve the semiclassical limit discussed in section \ref{subsec: semiclassical}, and attempt to study the duality in a healthier regime where the representations of the $\cW$-algebra are unitary.

\acknowledgments

We are grateful to our friend Juan Jottar, without whom this work would not have been possible. In fact, most of the heavy lifting was done by Juan in 2015 while working on our companion paper \cite{Banados:2015tft}, and we thank him for making his notes and Mathematica code available to us. We also thank Max Ba\~nados and Jan de Boer for useful discussions. A.C. is supported by the Delta ITP consortium, a program of the NWO that is funded by the Dutch Ministry of Education, Culture and Science (OCW). The work of A.F. was supported by CONICYT FONDECYT Regular \#1201145 and \#1160282. The work of I.O. was supported by CONICYT FONDECYT Regular \#1160282.

\appendix
\section{Composite operators}\label{app: quantum composites}
In this appendix we display the full quantum composite operators $\cA{s}$, $\cB{s}$, $\cC{s}$, $\cD{s}$, $\cE{s}{\pm}$, $\cPhi{s}{\pm}$, $\cPsi{s}{\pm}$ appearing in the $\cW_{(3|2)}$ algebra together with their transformation properties under the spectral flow automorphism.
\subsection{Definition}\label{app: quantum composites def}
In what follows, all combinations of fields enclosed by square brackets are quasi-primary. Naturally, normal-ordering is assumed. As in \cite{Romans:1991wi} we define
\begin{align}
	\gamma&\equiv\frac{1}{\left(c-1\right)\left(c+6\right)\left(2c-3\right)}\,,
	&
	\kappa&\equiv\pm\frac{(c+3)(5c-12)}{\sqrt{2(c+6)(c-1)(2c-3)(15-c)}}\,.
\end{align}
The composite operators appearing in the $\cW_{(3|2)}$ algebra then read
\begin{flalign*}
	\numberthis
	\label{quantum composites start}
	\cA{2}&=\frac{c}{c-1}\left(\left[T\right]-\frac{3}{2c}\left[J^{2}\right]\right)+\kappa\left[V\right]\,,&
	\\
	\numberthis
	\cB{2}&=\frac{1}{20\left(c-1\right)}\left(\left(5c-4\right)\left[T\right]-\frac{3}{2}\left[J^2\right]\right)+\frac{\kappa}{20}\left[V\right]\,,&
	\\
	\numberthis
	\cB{4}&=3\gamma\Bigg\{\left(16c^2-27c+18\right)\left[T^2-\frac{3}{10}\partial^2T\right]
	\\
	&+\frac{7}{2}c\left(c-6\right)\left[\partial G^+G^--G^+\partial G^-+\frac{2}{5}\partial^2T+\frac{1}{6}\partial^3J\right]
	\\
	&+6\left(4c+3\right)\left(\left[JG^+G^--J\partial T-\frac{1}{3}J\partial^2J\right]-\left[J^2T\right]\right)
	\\
	&+\frac{1}{4}\left(12c^2-61c+42\right)\left[J\partial^2J-\frac{3}{10}\partial^2J^2\right]\Bigg\}
	\\
	&+\frac{3\kappa}{\left(c+3\right)\left(5c-12\right)}\Bigg\{4\left(4c+3\right)\left[TV-\frac{3}{10}\partial^2V\right]
	\\
	&+\frac{1}{2}\left(11c+24\right)\left[G^-U^+-G^+U^-+\frac{2}{5}\partial^2V\right]+6\left(c+6\right)\left[JW\right]\Bigg\}\,,
	\\
	\numberthis\label{quantum composite C1}
	\cC{1}&=\textcolor{red}{\frac{1}{6}}\times\frac{1}{2}\left[J\right]\,,&
	\\
	\numberthis
	\cC{3}&=\gamma\Bigg\{\frac{3}{2}c\left(5c-12\right)\left[G^+G^--\partial T-\frac{1}{3}\partial^2J\right]+\left(8c^2-9c+36\right)\left[JT\right]
	\\
	&-3\left(4c+3\right)\left[J^3\right]\Bigg\}+\frac{\kappa}{5c-12}\Bigg\{\left(c-8\right)\left[W\right]+14\left[JV\right]\Bigg\}\,,
	\\
	\numberthis
	\cC{4}&=\frac{2}{c-1}\left[J\partial T-2\partial J T\right]+\frac{\kappa}{c+3}\Bigg\{2\left[J\partial V-2\partial JV\right]
	\\
	&-3\left[G^+U^-+G^-U^+-\frac{4}{3}\partial W\right]\Bigg\}\,,
	\\
	\numberthis
	\cD{1}&=\frac{1}{4}\left[J\right]\,,&
	\\
	\numberthis
	\cD{2}&=\frac{1}{10\left(c-1\right)}\left(\left(5c-3\right)\left[T\right]-3\left[J^2\right]\right)+\frac{\kappa}{5}\left[V\right]\,,
	\\
	\numberthis
	\cD{3}&=3\gamma\Bigg\{2\left(5c^2+9\right)\left[JT\right]-3\left(4c+3\right)\left[J^3\right]
	\\
	&+\frac{1}{2}\left(c-3\right)\left(13c-6\right)\left[G^+G^--\partial T-\frac{1}{3}\partial^2J\right]\Bigg\}+\frac{2\kappa}{5c-12}\Bigg\{21\left[JV\right]-\left(c+6\right)\left[W\right]\Bigg\}\,,
	\\
	\numberthis
	\cD{4}&=6\gamma\Bigg\{9c\left(c-1\right)\left[T^2-\frac{3}{10}\partial^2T\right]
	\\
	&+\frac{1}{4}\left(5c^2-51c+18\right)\left[\partial G^+G^--G^+\partial G^-+\frac{2}{5}\partial^2T+\frac{1}{6}\partial^3J\right]
	\\
	&+3\left(4c+3\right)\left(\left[JG^+G^--J\partial T-\frac{1}{3}J\partial^2J\right]-\left[J^2T\right]\right)
	\\
	&+\frac{1}{4}\left(c^2-53c+66\right)\left[J\partial^2J-\frac{3}{10}\partial^2J^2\right]\Bigg\}
	\\
	&+\frac{6\kappa}{\left(c+3\right)\left(5c-12\right)}\Bigg\{18\left(c-1\right)\left[TV-\frac{3}{10}\partial^2V\right]+\left(4c+3\right)\left[G^-U^+-G^+U^-+\frac{2}{5}\partial^2V\right]
	\\
	&-2\left(c-15\right)\left[JW\right]\Bigg\}\,,
	\\
	\numberthis
	\cE{4}{\pm}&=-\frac{6}{c-1}\left[\partial G^{\pm}G^{\pm}\right]\mp\frac{12\kappa}{c+3}\left[G^{\pm}U^{\pm}\right]\,,&
	\\
	\numberthis
	\cPhi{3/2}{\pm}&=\pm\frac{1}{4}\left[G^{\pm}\right]\,,&
	\\
	\numberthis
	\cPhi{5/2}{\pm}&=-\frac{6}{5\left(c-1\right)}\left[JG^{\pm}\mp\frac{1}{3}\partial G^{\pm}\right]+\frac{2\kappa}{5}\left[U^{\pm}\right]\,,
	\\
	\numberthis\label{quantum composite Phi7/2}
	\cPhi{7/2}{\pm}&=\textcolor{red}{2}\times\frac{3\gamma}{2}\Bigg\{\pm9c\left(c-1\right)\left[TG^{\pm}-\frac{3}{8}\partial^2G^{\pm}\right]-3\left(4c+3\right)\left[\pm J^2G^{\pm}-\partial JG^{\pm}\right]
	\\
	&-\frac{1}{10}\left(c^2-93c+36\right)\left[2J\partial G^{\pm}-3\partial JG^{\pm}\mp\frac{1}{4}\partial^2G^{\pm}\right]\Bigg\}
	\\
	&+\frac{6\kappa}{\left(c+3\right)\left(5c-12\right)}\Bigg\{9\left(c-1\right)\left[\pm VG^{\pm}-\frac{3}{5}\partial U^{\pm}\right]-\left(c-15\right)\left[\pm JU^{\pm}-\frac{1}{5}\partial U^{\pm}\right]\Bigg\}\,,
	\\
	\numberthis
	\cPsi{3/2}{\pm}&=\frac{1}{8}\left[G^{\pm}\right]\,,&
	\\
	\numberthis
	\cPsi{5/2}{\pm}&=-\frac{3}{10\left(c-1\right)}\left[\pm JG^{\pm}-\frac{1}{3}\partial G^{\pm}\right]\pm\frac{\kappa}{10}\left[U^{\pm}\right]\,,
	\\
	\numberthis
	\cPsi{7/2}{\pm}&=\frac{3\gamma}{7}\Bigg\{\left(55c^2-99c+72\right)\left[TG^{\pm}-\frac{3}{8}\partial^2G^{\pm}\right]-21\left(4c+3\right)\left[J^2G^{\pm}\mp\partial JG^{\pm}\right]
	\\
	&-\frac{1}{10}\left(47c^2-471c-108\right)\left[\pm2J\partial G^{\pm}\mp3\partial JG^{\pm}-\frac{1}{4}\partial^2G^{\pm}\right]\Bigg\}
	\\
	&+\frac{6\kappa}{7\left(c+3\right)\left(5c-12\right)}\Bigg\{\left(23c+33\right)\left[VG^{\pm}\mp\frac{3}{5}\partial U^{\pm}\right]+\left(13c+57\right)\left[JU^{\pm}\mp\frac{1}{5}\partial U^{\pm}\right]\Bigg\}\,,
	\\
	\numberthis
	\label{quantum composites finish}
	\cPsi{9/2}{\pm}&=\frac{3}{7\left(c-1\right)}\Bigg\{2\left[3\partial TG^{\pm}-4T\partial G^{\pm}+\frac{4}{5}\partial^3G^{\pm}\right]
	\\
	&-\left[\pm2\partial^2JG^{\pm}\mp4\partial J\partial G^{\pm}\pm J\partial^2G^{\pm}-\frac{1}{15}\partial^3G^{\pm}\right]\Bigg\}
	\\
	&+\frac{3\kappa}{7\left(c+3\right)}\Bigg\{\pm14\left[TU^{\pm}-\frac{1}{4}\partial^2U^{\pm}\right]\mp14\left[WG^{\pm}-\frac{5}{6}\partial^2U^{\pm}\right]
	\\
	&+\left[3\partial VG^{\pm}-4V\partial G^{\pm}\pm\partial^2U^{\pm}\right]+\left[2J\partial U^{\pm}-5\partial JU^{\pm}\mp\frac{1}{6}\partial^2U^{\pm}\right]\Bigg\}\,.
\end{flalign*}
We have highlighted in red the coefficients in $\cC{1}$ and $\cPhi{7/2}{\pm}$ that need to be modified with respect to \cite{Romans:1991wi} so that the Jacobi identities are satisfied and the algebra is spectral flow-invariant.
\subsection{Spectral flow}\label{app: quantum composites spectral flow}
In order to derive the spectral flow rules for the composite operators we need to recall the definition of normal-ordering as the regular part of the OPE in the coincidence limit. Following the conventions of \cite{Romans:1991wi,Thielemans:1994er} such that the OPE between two holomorphic operators $A$ and $B$ is written as
\begin{empheq}{alignat=7}
	A(z)B(w)&=\sum_{l\leq h}\frac{[AB]_l(w)}{(z-w)^l}\,,
\end{empheq}
with $h\leq h_A+h_B$, the normal-ordering prescription is
\begin{empheq}{alignat=7}
	{:}AB{:}(w)&\equiv\lim_{z\to w}\left(A(z)B(w)-\sum_{l>0}\frac{[AB]_l(w)}{(z-w)^l}\right)=[AB]_0(w)\,.
\end{empheq}
Products of more than two fields are defined recursively, grouping them as
\begin{empheq}{alignat*=7}
	:A_1A_2\cdots A_i:&=\left(:A_1\left(:A_2\left(\cdots\left(:A_{i-1}A_i:\right)\cdots\right):\right):\right)\,.
\end{empheq}
These definitions, together with the OPE algebra, allow us to compute the spectral flow of any composite starting from the transformation rules for the fundamental fields.

Take as an example the composite ${:}G^+G^-{:}(w)$, entering in the definition of $\cC{3}$ and $\cD{3}$. Using \eqref{spectral flow currents W1}, we first compute the OPE between the spectral flowed operators, which in the above notation becomes
\begin{empheq}{alignat=7}
	G^{+'}(z)G^{-'}(w)&=z^{\eta}w^{-\eta}G^+(z)G^-(w)=z^{\eta}w^{-\eta}\sum_{l=-\infty}^3\frac{[G^+G^-]_l(w)}{(z-w)^l}\,.
\end{empheq}
Since the $z$ dependence on the right hand side can only involve powers of $z-w$, we expand
\begin{empheq}{alignat=7}\label{z expansion}
	z^{\eta}&=w^{\eta}\sum_{k=0}^{\infty}\binom{\eta}{k}\left(\frac{z-w}{w}\right)^k\,.
\end{empheq}
Relabeling the indices and swapping the sums we find
\begin{empheq}{alignat=7}
	G^{+'}(z)G^{-'}(w)&=G^+(z)G^-(w)+\sum_{l=-\infty}^{3}\sum_{k=1}^{3-l}\binom{\eta}{k}\frac{w^{-k}[G^+G^-]_{l+k}(w)}{(z-w)^l}\,.
\end{empheq}
The new OPE can be read directly from this expression. In particular, the regular part is
\begin{empheq}{alignat=7}
	{:}G^{+'}G^{-'}{:}(w)&={:}G^+G^-{:}(w)+\sum_{k=1}^{3}\binom{\eta}{k}w^{-k}[G^+G^-]_l(w)\,,
\end{empheq}
which, after extracting
\begin{empheq}{alignat=7}
	[G^+G^-]_3(w)&=\frac{2c}{3}\,,
	&\qquad
	[G^+G^-]_2(w)&=2J(w)\,,
	&\qquad
	[G^+G^-]_1(w)&=2T(w)+\partial J(w)\,,
\end{empheq}
from \eqref{Super Virasoro finish}, reads
\begin{equation}
	{:}G^{+'}G^{-'}{:}(w)={:}G^+G^-{:}(w)+\frac{\eta}{w}\left(2T(w)+\partial J(w)\right)+\frac{\eta(\eta-1)}{w^2}J(w)+\frac{c\eta(\eta-1)(\eta-2)}{9w^3}\,.
\end{equation}
It also follows from the singular part that
\begin{empheq}{alignat=7}
	G^{+'}(z)G^{-'}(w)&\sim\frac{2c/3}{(z-w)^3}+\frac{2J'(w)}{(z-w)^2}+\frac{2T'(w)+\partial J'(w)}{(z-w)}\,,
\end{empheq}
showing the invariance of this particular OPE.

The procedure is similar in all other cases, always relying on the OPE algebra and the expansion \eqref{z expansion} for different values of $\eta$. This way, starting from the definitions \eqref{quantum composites start}-\eqref{quantum composites finish} of the quantum composites and using the transformation rules \eqref{spectral flow currents W1} and \eqref{spectral flow currents W2} for the fundamental fields, we find that the spectral flowed-version of these operators is
\begin{flalign}
	\label{spectral flow composites start}
	\cA{2}\hspace{0.01cm}'&=\cA{2}\,,&
	\\
	\cB{2}\hspace{0.01cm}'&=\cB{2}+\frac{12\eta}{5z}\cC{1}+\frac{c\eta^2}{30z^2}\,,
	\\
	\cB{4}\hspace{0.01cm}'&=\cB{4}+\frac{6\eta}{z}\cC{3}+\frac{36\eta}{5z}\left(\partial^{2}+\frac{3}{z}\partial+\frac{2}{z^2}\right)\cC{1}+\frac{4\eta^2}{z^2}\cA{2}+\frac{c\eta^2}{10z^4}\,,
	\\\label{spectral flow C1}
	\cC{1}\hspace{0.01cm}'&=\cC{1}+\frac{c\eta}{36z}\,,
	\\\label{spectral flow C3}
	\cC{3}\hspace{0.01cm}'&=\cC{3}+\frac{4\eta}{3z}\cA{2}\,,
	\\\label{spectral flow C4}
	\cC{4}\hspace{0.01cm}'&=\cC{4}+\frac{2\eta}{3z}\left(\partial+\frac{2}{z}\right)\cA{2}\,,
	\\
	\cD{1}\hspace{0.01cm}'&=\cD{1}+\frac{c\eta}{12z}\,,
	\\
	\cD{2}\hspace{0.01cm}'&=\cD{2}+\frac{6\eta}{5z}\cD{1}+\frac{c\eta^2}{20z^2}\,,
	\\
	\cD{3}\hspace{0.01cm}'&=\cD{3}+\frac{10\eta}{z}\cD{2}+\frac{6\eta^2}{z^2}\cD{1}+\frac{c\eta^3}{6z^3}\,,
	\\
	\nonumber
	\cD{4}\hspace{0.01cm}'&=\cD{4}+\frac{2\eta}{z}\cD{3}+\frac{10\eta^2}{z^2}\cD{2}+\frac{2\eta}{5z}\left(\partial^{2}+\frac{3}{z}\partial+\frac{2(1+5\eta^2)}{z^2}\right)\cD{1}
	\\
	&+\frac{c\eta^2(1+5\eta^2)}{60w^4}\,,
	\\
	\cE{4}{\pm}\hspace{0.01cm}'&=z^{\pm 2\eta}\cE{4}{\pm}\,,
	\\
	\cPhi{3/2}{\pm}\hspace{0.01cm}'&=z^{\pm\eta}\cPhi{3/2}{\pm}\,,
	\\
	\cPhi{5/2}{\pm}\hspace{0.01cm}'&=z^{\pm\eta}\left[\cPhi{5/2}{\pm}\mp\frac{8\eta}{5z}\cPhi{3/2}{\pm}\right]\,,
	\\
	\cPhi{7/2}{\pm}\hspace{0.01cm}'&=z^{\pm\eta}\left[\cPhi{7/2}{\pm}\mp\frac{\eta}{z}\cPhi{5/2}{\pm}\pm\frac{\eta}{5z}\left(-2\partial+\frac{3}{z}(\pm\eta-1)\right)\cPhi{3/2}{\pm}\right]\,,
	\\
	\cPsi{3/2}{\pm}\hspace{0.01cm}'&=z^{\pm\eta}\cPsi{3/2}{\pm}\,,
	\\
	\cPsi{5/2}{\pm}\hspace{0.01cm}'&=z^{\pm\eta}\left[\cPsi{5/2}{\pm}\pm\frac{4\eta}{z}\cPsi{3/2}{\pm}-\frac{12\eta}{5z}\cPhi{3/2}{\pm}\right]\,,
	\\
	\nonumber
	\cPsi{7/2}{\pm}\hspace{0.01cm}'&=z^{\pm\eta}\left[\cPsi{7/2}{\pm}\pm\frac{12\eta}{7z}\cPsi{5/2}{\pm}+\frac{6\eta}{35z}\left(\pm22\partial+\frac{47}{z}(\eta\mp 1)\right)\cPsi{3/2}{\pm}\right.
	\\
	&\left.+\frac{10\eta}{7z}\cPhi{5/2}{\pm}-\frac{2\eta}{7z}\left(16\partial\pm24\frac{\eta}{z}\right)\cPhi{3/2}{\pm}\right]\,,
	\\
	\nonumber
	\cPsi{9/2}{\pm}\hspace{0.01cm}'&=z^{\pm\eta}\left[\cPsi{9/2}{\pm}\pm\frac{2\eta}{z}\cPsi{7/2}{\pm}\mp\frac{2\eta}{7z}\left(2\partial\mp\frac{5}{z}(\eta\mp 1)\right)\cPsi{5/2}{\pm}\right.
	\\
	\nonumber
	&-\frac{4\eta}{35z}\left(\pm 67\partial^{2}+\frac{164}{z}(\eta\mp 1)\partial\pm\frac{62}{z^2}(\eta\mp 2)(\eta\mp 1)\right)\cPsi{3/2}{\pm}-\frac{2\eta}{z}\cPhi{7/2}{\pm}
	\\
	\nonumber
	&+\frac{\eta}{7z}\left(6\partial\pm\frac{5}{z}(4\eta\pm3)\right)\cPhi{5/2}{\pm}+\frac{2\eta}{35z}\left(57\partial^{2}\pm\frac{4}{z}(26\eta-51)\partial\right.
	\\
	\label{spectral flow composites finish}
	&\left.\left.+\frac{12}{z^2}(\eta^2\mp 23\eta+7)\right)\cPhi{3/2}{\pm}\right]\,.
\end{flalign}
We emphasize that the above rules were not derived by demanding that the full $\cW_{(3|2)}$ algebra be spectral flow-invariant. Instead, we have verified this fact a posteriori using the Mathematica package of \cite{Thielemans:1994er}. It is instructive, however, to follow the reverse process and deduce the transformation properties of the composites $\cC{1}$, $\cC{3}$ and $\cC{4}$ by requiring the invariance of the $VW$ OPE. We choose this particular example because it is simple and involves the composite $\cC{1}$, which we claim needs to be corrected with respect to \cite{Romans:1991wi}. Using \eqref{spectral flow currents W2} and \eqref{VW OPE} we get
\begin{flalign}
	\numberthis
	V'(z)W'(w)&\sim\frac{\cC{4}(w)}{z-w}+\left(\frac{3}{(z-w)^{2}}+\frac{1}{z-w}\partial \right)\cC{3}(w)+\frac{36}{(z-w)^{4}}\cC{1}(w)
	\\
	&+\frac{2\eta}{w}\left(\frac{c/2}{(z-w)^{4}}+\left(\frac{2}{(z-w)^{2}}+\frac{1}{z-w}\partial\right)\cA{2}(w)\right)\,.
\end{flalign}
Collecting poles of same order we find that spectral flow invariance demands that
\begin{empheq}{alignat=7}
	\cC{4}\hspace{0.01cm}'(w)+\partial\cC{3}\hspace{0.01cm}'(w)&=\cC{4}(w)+\partial\cC{3}(w)+\frac{2\eta}{w}\partial\cA{2}(w)\,,
	\\
	\cC{3}\hspace{0.01cm}'(w)&=\cC{3}(w)+\frac{4\eta}{3w}\cA{2}(w)\,,
	\\
	\cC{1}\hspace{0.01cm}'(w)&=\cC{1}(w)+\frac{c\eta}{36w}\,.
\end{empheq}
After substituting the second equation into the first one, these rules agree with \eqref{spectral flow C1}, \eqref{spectral flow C3} and \eqref{spectral flow C4}. Notice that the transformation property of $\cC{1}$ follows from that of $J$ only if we correct the coefficient in \eqref{quantum composite C1}. A similar approach can be taken to fix \eqref{quantum composite Phi7/2}.

\section{The $sl(3|2)$ superalgebra}\label{app: sl(3|2)}
In this appendix we collect some useful facts and formulae regarding the superalgebra $sl(3|2)$ and its real form $su(2,1|1,1)\,$.
\subsection{Definition and (anti-)commutation relations}\label{app: sl(3|2) definition and commutators} 
The superalgebra $sl(m|n;\mathds{C})$ consists of all complex $(m+n)\times(m+n)$ supermatrices of the form
\begin{empheq}{alignat=5}
	M&=\left(
	\begin{array}{c|c}
		A & B \\\hline
		C & D
	\end{array}
	\right)\,,
\end{empheq}
equipped with the supercommutator
\begin{empheq}{align}
	[M,M'\}&=\left(
	\begin{array}{c|c}
		AA'-A'A+BC'+B'C & AB'-A'B+BD'-B'D \\\hline
		CA'-C'A+DC'-D'C & CB'+C'B+DD'-D'D
	\end{array}
	\right)\,,
\end{empheq}
and satisfying the supertraceless condition
\begin{empheq}{alignat=5}
	\textrm{sTr}(M)&\equiv\textrm{Tr}\left[A\right]-\textrm{Tr}\left[D\right]&\,=0\,.
\end{empheq}
The complex dimension of the superalgebra is $(m+n)^2-1$. Elements with $B=0$ and $C=0$ are called even or bosonic, while those with $A=0$ and $D=0$ are termed odd or fermionic. The even subalgebra is $sl(m;\mathds{C})\oplus sl(n;\mathds{C})\oplus\mathds{C}$. In what follows we deal specifically with $m=3$ and $n=2$. We comment on the real form of interest below.

In the principal embedding of $sl(2|1)$ in $sl(3|2)$ \cite{Peng:2012ae,Chen:2013oxa}, the even-graded sector of the superalgebra is decomposed into the $sl(2)$ generators, $L_i$, one spin 1 multiplet, $A_i$, one spin 2 multiplet, $W_m$, and a spin 0 element, $J$. By spin we mean the $sl(2)$ spin, $S$. Within each multiplet the indices range from $-S$ to $S$, giving a total of $3+3+5+1=12$ bosonic generators. This structure is encoded in the commutation relations
\begin{empheq}{alignat=5}
	[L_i,L_j]&=(i-j)L_{i+j}\,,
	&\qquad
	[L_i,A_j]&=(i-j)A_{i+j}\,,
	&\qquad
	[L_i,W_m]&=(2i-m)W_{i+m}\,.
\end{empheq}
The remaining non-vanishing commutators read
\begin{empheq}{alignat=5}
	[A_i,A_j]&=(i-j)L_{i+j}\,,
	\qquad
	[A_i,W_m]=(2i-m)W_{i+m}\,,
	\\\nonumber
	[W_m,W_n]&=-\frac{1}{6}(m-n)(2m^2+2n^2-mn-8)(L_{m+n}+A_{m+n})\,.
\end{empheq}
Therefore, the bosonic part of the $sl(3|2)$ algebra is $sl(3)\oplus sl(2)\oplus u(1)$, where the $sl(3)$ is generated by $(L_i+A_i)/2$ together with $W_m$, while the $sl(2)$ corresponds to $(L_i-A_i)/2$. The latter factor should not be confused with the ``gravitational'' $sl(2)$ spanned by $L_i$. Of course, the Abelian generator is $J$. In turn, the odd-graded elements consist of two spin $1/2$ multiplets, $H_r$ and $G_r$, and two spin $3/2$ multiplets, $T_s$ and $S_s$;
\begin{empheq}{alignat=5}
	[L_i,G_r]&=\left(\frac{i}{2}-r\right)G_{i+r}\,,
	&\qquad
	[L_i,H_r]&=\left(\frac{i}{2}-r\right)H_{i+r}\,,
	\\\nonumber
	[L_i,S_s]&=\left(\frac{3i}{2}-s\right)S_{i+s}\,,
	&\qquad
	[L_i,T_s]&=\left(\frac{3i}{2}-s\right)T_{i+s}\,.
\end{empheq}
The number of fermionic generators is $2+2+4+4=12$. Their $U(1)$ charge assignments are
\begin{empheq}{alignat=5}
	[J,G_r]&=G_r\,,
	&\qquad
	[J,H_r]&=-H_r\,,
	&\qquad
	[J,S_s]&=S_s\,,
	&\qquad
	[J,T_s]&=-T_s\,.
\end{empheq}
Additionally, they satisfy
\begin{empheq}{alignat=5}
	[A_i,G_r]&=\frac{5}{3}\left(\frac{i}{2}-r\right)G_{i+r}+\frac{4}{3}S_{i+r}\,,
	\qquad
	[A_i,H_r]=\frac{5}{3}\left(\frac{i}{2}-r\right)H_{i+r}-\frac{4}{3}T_{i+r}\,,
	\\\nonumber
	[A_i,S_s]&=\frac{1}{3}\left(\frac{3i}{2}-s\right)S_{i+s}-\frac{1}{3}\left(3i^2-2is+s^2-\frac{9}{4}\right)G_{i+s}\,,
	\\\nonumber
	[A_i,T_s]&=\frac{1}{3}\left(\frac{3i}{2}-s\right)T_{i+s}+\frac{1}{3}\left(3i^2-2is+s^2-\frac{9}{4}\right)H_{i+s}\,,
\end{empheq}
\begin{empheq}{alignat=5}
	[W_m,G_r]&=-\frac{4}{3}\left(\frac{m}{2}-2r\right)S_{m+r}\,,
	\qquad
	[W_m,H_r]=-\frac{4}{3}\left(\frac{m}{2}-2r\right)T_{m+r}\,,
	\\\nonumber
	[W_m,S_s]&=-\frac{1}{3}\left(2s^2-2sm+m^2-\frac{5}{2}\right)S_{m+s}
	\\\nonumber
	&-\frac{1}{6}\left(4s^3-3s^2m+2sm^2-m^3-9s+\frac{19}{4}m\right)G_{m+s}\,,
	\\\nonumber
	[W_m,T_s]&=\frac{1}{3}\left(2s^2-2sm+m^2-\frac{5}{2}\right)T_{m+s}
	\\\nonumber
	&-\frac{1}{6}\left(4s^3-3s^2m+2sm^2-m^3-9s+\frac{19}{4}m\right)H_{m+s}\,,
\end{empheq}
together with the anti-commutation relations
\begin{empheq}{alignat=5}
	\{G_r,H_s\}&=2L_{r+s}+(r-s)J\,,
	\\\nonumber
	\{G_r,T_s\}&=-\frac{3}{2}W_{r+s}+\frac{3}{4}(3r-s)A_{r+s}-\frac{5}{4}(3r-s)L_{r+s}\,,
	\\\nonumber
	\{H_r,S_s\}&=-\frac{3}{2}W_{r+s}-\frac{3}{4}(3r-s)A_{r+s}+\frac{5}{4}(3r-s)L_{r+s}\,,
	\\\nonumber
	\{S_r,T_s\}&=-\frac{3}{4}(r-s)W_{r+s}+\frac{1}{8}\left(3s^2-4rs+3r^2-\frac{9}{2}\right)\left(L_{r+s}-3A_{r+s}\right)
	\\\nonumber
	&-\frac{1}{4}(r-s)\left(r^2+s^2-\frac{5}{2}\right)J\,.
\end{empheq}
Notice that the elements $L_i$, $J$, $H_r$ and $G_r$ generate $sl(2|1)\subset sl(3|2)$, while $osp(1|2)\subset sl(2|1)$ is spanned by $L_i$ and $(H_r+G_r)/\sqrt{2}$.
\subsection{Matrix representation}\label{app: sl(3|2) representation}
For convenience, we have chosen to work in a representation where all matrices are real and satisfy
\begin{empheq}{alignat=5}
	L_i^{\dagger}=(-1)^iL_{-i}\,,
	&\qquad
	A_i^{\dagger}=(-1)^iA_{-i}\,,
	&\qquad
	W_m^{\dagger}=(-1)^mW_{-m}\,,
\end{empheq}
and
\begin{empheq}{alignat=5}
	H_r^{\dagger}&=(-1)^{r+\frac{1}{2}}G_{-r}\,,
	&\qquad
	T_s^{\dagger}&=(-1)^{s+\frac{1}{2}}S_{-s}\,.
\end{empheq}
The generators in this basis are \cite{Chen:2013oxa}
\begin{empheq}{alignat=5}
	L_1&=\left(
	\begin{array}{ccc|cc}
		0 & 0 & 0 & 0 & 0 \\
		\sqrt{2} & 0 & 0 & 0 & 0 \\
		0 & \sqrt{2} & 0 & 0 & 0 \\\hline
		0 & 0 & 0 & 0 & 0 \\
		0 & 0 & 0 & 1 & 0 \\
	\end{array}
	\right)\,,
	&\qquad
	L_0&=\left(
	\begin{array}{ccccc}
		1 & 0 & 0 & 0 & 0 \\
		0 & 0 & 0 & 0 & 0 \\
		0 & 0 & -1 & 0 & 0 \\
		0 & 0 & 0 & \frac{1}{2} & 0 \\
		0 & 0 & 0 & 0 & -\frac{1}{2} \\
	\end{array}
	\right)\,,
\end{empheq}
\begin{empheq}{alignat=5}
	A_1&=\left(
	\begin{array}{ccccc}
		0 & 0 & 0 & 0 & 0 \\
		\sqrt{2} & 0 & 0 & 0 & 0 \\
		0 & \sqrt{2} & 0 & 0 & 0 \\
		0 & 0 & 0 & 0 & 0 \\
		0 & 0 & 0 & -1 & 0 \\
	\end{array}
	\right)\,,
	&\qquad
	A_0&=\left(
	\begin{array}{ccccc}
		1 & 0 & 0 & 0 & 0 \\
		0 & 0 & 0 & 0 & 0 \\
		0 & 0 & -1 & 0 & 0 \\
		0 & 0 & 0 & -\frac{1}{2} & 0 \\
		0 & 0 & 0 & 0 & \frac{1}{2} \\
	\end{array}
	\right)\,,
\end{empheq}
\begin{empheq}{alignat=5}
	W_2&=\left(
	\begin{array}{ccccc}
		0 & 0 & 0 & 0 & 0 \\
		0 & 0 & 0 & 0 & 0 \\
		4 & 0 & 0 & 0 & 0 \\
		0 & 0 & 0 & 0 & 0 \\
		0 & 0 & 0 & 0 & 0 \\
	\end{array}
	\right)\,,
	&\qquad
	W_1&=\left(
	\begin{array}{ccccc}
		0 & 0 & 0 & 0 & 0 \\
		\sqrt{2} & 0 & 0 & 0 & 0 \\
		0 & -\sqrt{2} & 0 & 0 & 0 \\
		0 & 0 & 0 & 0 & 0 \\
		0 & 0 & 0 & 0 & 0 \\
	\end{array}
	\right)\,,
\end{empheq}
\begin{empheq}{alignat=5}
	W_0&=\left(
	\begin{array}{ccccc}
		\frac{2}{3} & 0 & 0 & 0 & 0 \\
		0 & -\frac{4}{3} & 0 & 0 & 0 \\
		0 & 0 & \frac{2}{3} & 0 & 0 \\
		0 & 0 & 0 & 0 & 0 \\
		0 & 0 & 0 & 0 & 0 \\
	\end{array}
	\right)\,,
	&\qquad
	J&=\left(
	\begin{array}{ccccc}
		2 & 0 & 0 & 0 & 0 \\
		0 & 2 & 0 & 0 & 0 \\
		0 & 0 & 2 & 0 & 0 \\
		0 & 0 & 0 & 3 & 0 \\
		0 & 0 & 0 & 0 & 3 \\
	\end{array}
	\right)\,,
\end{empheq}
\begin{empheq}{alignat=5}
	G_{\frac{1}{2}}&=\left(
	\begin{array}{ccc|cc}
		0 & 0 & 0 & 0 & 0 \\
		0 & 0 & 0 & 0 & 0 \\
		0 & 0 & 0 & 0 & 0 \\\hline
		2 & 0 & 0 & 0 & 0 \\
		0 & \sqrt{2} & 0 & 0 & 0 \\
	\end{array}
	\right)\,,
	&\qquad
	H_{\frac{1}{2}}&=\left(
	\begin{array}{ccccc}
		0 & 0 & 0 & 0 & 0 \\
		0 & 0 & 0 & \sqrt{2} & 0 \\
		0 & 0 & 0 & 0 & 2 \\
		0 & 0 & 0 & 0 & 0 \\
		0 & 0 & 0 & 0 & 0 \\
	\end{array}
	\right)\,,
\end{empheq}
\begin{empheq}{alignat=5}
	S_{\frac{3}{2}}&=\left(
	\begin{array}{ccccc}
		0 & 0 & 0 & 0 & 0 \\
		0 & 0 & 0 & 0 & 0 \\
		0 & 0 & 0 & 0 & 0 \\
		0 & 0 & 0 & 0 & 0 \\
		-3 & 0 & 0 & 0 & 0 \\
	\end{array}
	\right)\,,
	&\qquad
	S_{\frac{1}{2}}&=\left(
	\begin{array}{ccccc}
		0 & 0 & 0 & 0 & 0 \\
		0 & 0 & 0 & 0 & 0 \\
		0 & 0 & 0 & 0 & 0 \\
		-1 & 0 & 0 & 0 & 0 \\
		0 & \sqrt{2} & 0 & 0 & 0 \\
	\end{array}
	\right)\,,
\end{empheq}
\begin{empheq}{alignat=5}
	T_{\frac{3}{2}}&=\left(
	\begin{array}{ccccc}
		0 & 0 & 0 & 0 & 0 \\
		0 & 0 & 0 & 0 & 0 \\
		0 & 0 & 0 & -3 & 0 \\
		0 & 0 & 0 & 0 & 0 \\
		0 & 0 & 0 & 0 & 0 \\
	\end{array}
	\right)\,,
	&\qquad
	T_{\frac{1}{2}}&=\left(
	\begin{array}{ccccc}
		0 & 0 & 0 & 0 & 0 \\
		0 & 0 & 0 & -\sqrt{2} & 0 \\
		0 & 0 & 0 & 0 & 1 \\
		0 & 0 & 0 & 0 & 0 \\
		0 & 0 & 0 & 0 & 0 \\
	\end{array}
	\right)\,.
\end{empheq}
\subsection{The real form $su(2,1|1,1)$}\label{app: sl(3|2) real form} 
The superalgebra $su(2,1|1,1)\supset su(2,1)\oplus su(1,1)\oplus i\mathds{R}$ is defined as the set of supertraceless $5\times5$ supermatrices $M$ satisfying
\begin{empheq}{alignat=5}
	M^{\dagger}K+KM&=0\,,
\end{empheq}
where $K$ is a non-degenerate Hermitian form of signature $(2,1|1,1)$. One can check that in our representation of $sl(3|2)$ the generators
\begin{empheq}{alignat=5}
	L_i,\quad A_i,\quad iW_m,\quad iJ,\,
\end{empheq}
and
\begin{empheq}{alignat=5}
	e^{i\pi/4}\left(H_r+G_r\right),\quad e^{-i\pi/4}\left(H_r-G_r\right),\quad e^{-i\pi/4}\left(T_s+S_s\right),\quad e^{i\pi/4}\left(T_s-S_s\right)\,,
\end{empheq}
satisfy the above property with
\begin{empheq}{alignat=5}
	K&=\left(
	\begin{array}{ccccc}
		0 & 0 & -1 & 0 & 0 \\
		0 & 1 & 0 & 0 & 0 \\
		-1 & 0 & 0 & 0 & 0 \\
		0 & 0 & 0 & 0 & i \\
		0 & 0 & 0 & -i & 0 \\
	\end{array}
	\right)\,.
\end{empheq}
Notice that $K$ has the correct eigenvalues. Therefore, these particular combinations of generators, with the above pre-factors included, form a basis for the \emph{real} superalgebra $su(2,1|1,1)$.
\bibliographystyle{JHEP}
\bibliography{W32}
\end{document}